\documentclass[12pt,preprint]{aastex}
\shorttitle{Exact Solutions of the Isothermal Lane--Emden Equation with Rotation}
\shortauthors{Christodoulou \& Kazanas}

\begin{document}

\title{ Exact Solutions \\ of the Isothermal Lane--Emden Equation with Rotation \\
        and Implications for the Formation of Planets and Satellites}

\author{
Dimitris M. Christodoulou\altaffilmark{1,2} 
and 
Demosthenes Kazanas\altaffilmark{3}
}
%\affil{}

\altaffiltext{1}{Math Methods, 54 Middlesex Tpke, Bedford, MA 01730.
			E-mail: dimitris@mathmethods.com}
\altaffiltext{2}{University of Massachusetts Lowell, Dept. of Mathematical
                 Sciences, Olney Hall, Room 428, \\ Lowell, MA 01854.
                        E-mail: Dimitris\_\_Christodoulou@uml.edu}
\altaffiltext{3}{NASA/GSFC, Code 663, Greenbelt, MD 20771.
			E-mail: Demos.Kazanas@nasa.gov}

\def\gsim{\mathrel{\raise.5ex\hbox{$>$}\mkern-14mu
             \lower0.6ex\hbox{$\sim$}}}

\def\lsim{\mathrel{\raise.3ex\hbox{$<$}\mkern-14mu
             \lower0.6ex\hbox{$\sim$}}}

\begin{abstract}

We have derived exact solutions of the isothermal Lane--Emden equation with and 
without rotation 
in a cylindrical geometry. The corresponding hydrostatic equilibria are relevant 
to the dynamics of the solar nebula before and during the stages of planet and 
satellite formation. The nonrotating solution for the mass density is analytic, 
nonsingular, monotonically decreasing with radius, and it satisfies easily the usual 
physical boundary conditions at the center. When differential rotation is added to the 
Lane--Emden equation, a new class of exact solutions for the mass density 
appears. We have determined all of these solutions analytically as well. 
Within this class, solutions that are power laws or combinations of power laws
are not capable of satisfying the associated boundary--value problem, but they are 
nonetheless of profound importance because they constitute "baselines"
to which the actual solutions approach when the central boundary conditions 
are imposed. Numerical integrations that enforce such physical boundary
conditions show that the actual radial equilibrium density profiles emerge from
the center close to the nonrotating solution, but once they cross below the
corresponding baselines, they cease to be monotonic. The actual solutions
are forced to oscillate permanently about the baseline solutions
without ever settling onto them because the central boundary conditions 
strictly prohibit the matching of the two types of solutions.

This oscillatory behavior of the isothermal solutions to the Lane--Emden 
boundary--value problem is entirely generic and extends to polytropic models as well.
Based on our results, we expect that  quasistatically--evolving protoplanetary disks
should develop oscillatory radial density profiles in their midplanes during 
the isothermal phase of their collapse. The peaks in these profiles correspond
to local gravitational potential minima and their radial locations are ideal
sites for the formation of protoplanets; sites that can be accentuated
during infall of more matter from the 
still--collapsing cloud. Indeed, a straightforward application to our solar system
using an oscillatory solution derived from a differentially--rotating
baseline yields a highly accurate match (mean relative error $\approx 4\%$) 
between the radial density peaks of the model and the semimajor axes of all 
the major planets and the dwarf planets, 
provided that the mean density profile between 0.8 AU and 11.3 AU 
falls off with radius as $R^{-2.5}$ (corresponding to a mean surface density variation 
of $R^{-1.5}$ that is consistent with the profile determined empirically by Weidenschilling 
in 1977). We believe then that for the first time in over two centuries
we have a mathematically rigorous explanation of all planetary orbits
in our solar system and the physics that is responsible
for planet formation at radii that are not in the least random or arbitrary.

\end{abstract}




\section{Introduction}\label{intro}

After a research effort that spans almost ten years, we have been able to derive 
exact solutions of the nonlinear differential equations that describe the equilibrium 
structures of differentially--rotating, self--gravitating fluids with cylindrical symmetry. 
Our results are directly applicable to quasi--equilibrium 
configurations that may develop early in the evolution of protostellar and 
protoplanetary disks, and we expect that they will help us understand the 
physical conditions that prevail in such systems long before the onset of 
accretion processes, gas ionization, and nonaxisymmetric evolution. 
The nonrotating analogues of our equations
have been legendary in the literature; they are known as the nonlinear
Lane--Emden, Thomas--Fermi, and Emden--Fowler equations. Introduction of rotation to these
equations complicates matters considerably and very little analytical work
has been carried out to date in this most interesting case.

Our results are also relevant to a long--standing problem in the formation
of our solar system that has preoccupied professional astronomers and nonprofessionals
for over two centuries, namely the locations of planetary orbits and the observed
"order" in the present solar system. Naturally, we too have ended up considering
the famous Titius--Bode "law" of planetary distances but not in an effort to
discover some hidden physical principle or previously unknown "universal" 
underpinning; on the contrary, we saw this "law" as an adverse interpretation
to different and more complex patterns that we detected in the actual data
and that we describe in \S~\ref{work}.
We begin in \S\S~\ref{LE}--\ref{TB} with a brief overview of the research
that has been carried out by other researchers on these topics.
We also mention in \S~\ref{work} some of the history 
of our work that has finally tied up together all the pertinent issues.
We refer especially to our past unsuccessful attempts to resolve the
question of planetary distances because it is such setbacks 
that eventually pave the way for a successful conclusion of the research effort.

\subsection{The Lane--Emden Equation}\label{LE}

The Lane--Emden differential equation (Lane 1869--70; Emden 1907) 
describes the equilibria of nonrotating fluids in which internal pressure 
balances self-gravity. Spherically symmetric solutions of this equation came to 
the attention of astrophysicists when Chandrasekhar included them in his 1939
monograph "An Introduction to the Study of Stellar Structure," but interest
in such solutions continued to be largely academic because real stars rotate and
rotation destroys spherical symmetry and modifies their internal profiles and
physical characteristics. In the latter half of the twentieth century, the
isothermal solution, commonly referred to as the "singular isothermal sphere,"
and its nonsingular modifications found some interesting applications to the
structures of collisionless systems such as globular clusters and early-type 
galaxies (Binney \& Tremaine 1987; Rix et al. 1997);
to the structures of large--scale gaseous systems such as X--ray halos 
around elliptical galaxies (Fabbiano 1989) and clusters of galaxies 
(Sarazin 1988; Fabian 1991); and to simplified models of gravitational 
lenses (Kochanek 1995).

Emden's work also attracted the attention of physicists outside the field of
astrophysics (Fowler 1914; Thomas 1927; Fermi 1927) who studied generalized
polytropic forms of the Lane--Emden equation for specific polytropic indices $n$.
The extensive studies of Fowler (1914, 1930, 1931) produced some singular
solutions for $n = 3$ and established the so--called Emden--Fowler equation
in the literature, while the works of Thomas (1927) and Fermi (1927)
produced the so--called Thomas--Fermi equation, an important milestone
in atomic theory. At present, both of these equations continue to be 
the subjects of investigations by physicists and mathematicians alike. Physicists
are drawn to the Emden--Fowler equation because it appears in the kinetics
of Landau--Ginzburg critical phenomena (see the detailed account of Dixon \& 
Tuszy\'nski 1990) and in the kinetics of combustion (Frank--Kamenetskii 1955).
Mathematicians use these nonlinear equations as laboratories to study a wide
variety of properties in their solutions---positivity, uniqueness, singularities,
monotonic vs. oscillatory behavior, and bifurcations---usually without having 
analytic solutions at hand (e.g., Wong 1975; Lions 1982; see also Goenner \& Havas
[2000] for a comprehensive list of all known exact analytic solutions;
as well as Berkovich [1997] and Goenner [2001] for classifications of some 
solutions obtained from Lie--group symmetries).

The cylindrical form of the classical Lane--Emden equation differs from the 
spherical form in just one coefficient and both forms have been included as 
special cases in studies of the generalized Emden--Fowler equation (Horedt 1986; 
Dixon \& Tuszy\'nski 1990; Berkovich 1997; Goenner \& Havas 2000). 
But until now, the cylindrical
Lane--Emden equation has not attracted attention on its own merit---not even
in astrophysics where it can be used as a simplified (but nonlinear) model 
of a disk--like or cylindrical self--gravitating gas. The only analytic studies 
of the cylindrical form that we are aware of have been made by Jeans (1914) 
and Robe (1968) who
found a Bessel--function solution of the linear $n = 1$ case with uniform rotation
and by Stod\'olkiewicz (1963) and Ostriker (1964) who solved the nonrotating 
isothermal ($n\to\infty$) case. Ostriker (1964), in particular, also showed 
that the radius and the mass per unit length are finite in all models with
$0\leq n < \infty$ and that, in contrast to the singular isothermal sphere,
the nonrotating, nonsingular, isothermal cylinder has finite mass per unit length
in spite of its infinite radial extent.

After the work of Robe (1968), some numerical studies of rotating isothermal 
cylinders and disks (Hansen et al. 1976; Schmitz \& Ebert 1986; Narita et al. 1990) 
and rotating polytropic cylinders (Schneider \& Schmitz 1995) also found nonmonotonic
solutions for the equilibrium density akin to the Bessel--function oscillatory
solution of the $n=1$ polytrope in uniform rotation; but understanding this
behavior proved elusive and so the nonmonotonic density profiles were marginalized.
Schneider \& Schmitz (1995), in particular, adopted some differentially rotating 
profiles and obtained numerical solutions for the radial density profiles of models 
with negative polytropic indices ($n < -1$). These authors did not study the 
isothermal models of interest here, and they did not explain the unusual 
properties of their solutions. Nevertheless, their work is by far the most closely 
related study to our work and their polytropic results should be considered 
in conjunction with our results from isothermal models. 

\subsection{The Titius--Bode Algorithm}\label{TB}

The numerical algorithm called the Titius--Bode "law" has been known 
for 240 years (e.g., Nieto 1972; Lecar 1973; Danby 1988). 
It relies on an ad--hoc  
geometric progression to describe the positions of the planets in the 
solar system and works fairly well out to Uranus but no farther (Jaki 1972).
The same phenomenology has also been applied to the satellites of the 
gaseous giant planets (Neuh$\ddot{\rm a}$user \& Feitzinger 1986). 
Two modern brief reviews of the history along with criticisms of 
this "law" have been written by Graner \& Dubrulle (1994) 
and by Hayes \& Tremaine (1998). Currently, the general consensus is 
that a satisfactory physical basis has not been found for this numerical 
coincidence despite serious efforts by many researchers in over two centuries.
Furthermore, opinions differ on whether such a physical basis exists at all.

Apparently, many researchers still believe that the Titius--Bode
algorithm does have a physical foundation and continue to work on
this problem. The last decade of the twentieth century, in particular,
saw a resurgence of investigations targeting precisely two questions:
the origin of the "law" (Graner \& Dubrulle 1994;  Dubrulle \& Graner 1994;
Li, Zhang, \& Li 1995; Nottale, Schumacher, \& Gay 1997)
and its statistical robustness against the null hypothesis 
(Hayes \& Tremaine 1998).

Hayes \& Tremaine (1998) found some statistical evidence that the Titius--Bode
"law" could be related to the long--term dynamical stability of the solar system.
Their work is the latest in a long line of differing arguments made by
several statisticians in the past and summarized by these authors (see also Lynch 2003).
More importantly, Hayes \& Tremaine (1998) dismissed without discussion all the 
previous purported explanations of the physical origin of the "law" 
as "not entirely convincing." 
In our opinion, that is an example of good physical instincts, but it still 
requires some physical understanding of planetary distances before final judgment
is passed. We believe that the work presented below does provide the required 
understanding, and we will return to a broad discussion of the 
Titius--Bode "law" in \S~\ref{dis} below.

\subsection{Past Analyses of Planetary Distances and the Lane--Emden Equation}\label{work}

Just as Jeans (1914) had done years ago, we began studying rotating 
self--gravitating cylinders in 1997 for the same reason; as Jeans put it: 

\noindent
{\it "All the essential
physical features of the natural three--dimensional problem appear to be
reproduced in the simpler cylindrical problem, so that it seems legitimate
to hope that an argument by analogy may not lead to entirely erroneous
result[s]."} 

\noindent
Being unaware of the work published in French by Robe (1968),
we solved analytically the $n=1$ polytropic 
Lane--Emden equation with cylindrical symmetry, uniform rotation, and proper 
boundary conditions (Christodoulou 1997, unpublished). 
(Applying boundary conditions did not concern Jeans
who was interested in the stability of local deformations in compressible gases.)
The radial density profiles for $n=1$ and for sufficiently fast rotation exhibited
permanent oscillations due to the zeroth--order Bessel function $J_0(R)$ 
that is the dominant part of the solution. Numerical integrations for
polytropes with $n > 1$ also showed that the solutions for the density
oscillate permanently with radius. This generic oscillatory behavior 
then led us to consider the distribution of semimajor axes of planetary orbits 
in the solar system and their possible
connection to the radial density peaks found in the equilibrium solutions.

%---------------------------------------------------------------------
%                             TABLE 1
\begin{table}[t]
\begin{center}
\begin{minipage}{7in}
\caption{}
\begin{tabular}{cccccccccc}
\multicolumn{10}{c}{\sc Planetary Distances in Our Solar System} \\
\hline\hline
Index & Object & Semimajor  & Titius--Bode &       & Inversion &       & Mean     &       &             \\
 ~$i$ & Name   & Axis ~$a_i$& Distance     & Error & Distance  & Error & Distance & Error & ${\cal M}_i$\\
      &        &  (AU)      & (AU)         & (\%)  & (AU)      & (\%)  & (AU)     & (\%)  &             \\
\hline\hline
1  & Mercury & 0.387 & 0.4  & ~~3.4    &       &         &       &          &   \\
2  & Venus   & 0.723 & 0.7  & $-3.2$   & 0.622 & $-14.0$ & 0.694 & $-4.1$   & 1 \\
3  & Earth   & 1     & 1.0  & ~~0.0    & 1.050 & ~~5.0   & 1.124 & ~~12.4   & 2 \\
4  & Mars    & 1.524 & 1.6  & ~~5.0    & 1.663 & ~~9.1   & 1.883 & ~~23.5   & 2 \\
5  & Ceres   & 2.765 & 2.8  & ~~1.3    & 2.816 & ~~1.8   & 3.364 & ~~21.6   & 2 \\
6  & Jupiter & 5.203 & 5.2  & $-0.1$   & 5.135 & $-1.3$  & 6.151 & ~~18.2   & 2 \\
7  & Saturn  & 9.537 & 10.0 & ~~4.9    & 9.992 & ~~4.8   & 12.20 & ~~27.9   & 2 \\
8  & Uranus  & 19.19 & 19.6 & ~~2.1    & 16.93 & $-11.8$ & 19.80 & ~~3.2    & 1 \\
9  & Neptune & 30.07 & 38.8 & ~~29.0   & 27.52 & $-8.5$  & 29.34 & $-2.4$   & 1 \\
10 & Pluto   & 39.48 & 77.2 & ~~95.5   &       &         &       &          &   \\
\hline\hline
\multicolumn{10}{l}{{\sc Notes}.---Inversion Distance $= (a_{i-1}\cdot a_{i+1})^{1/2}$.
\ \ Mean Distance $ = \frac{1}{2}(a_{i-1} + a_{i+1})$.} \\
\multicolumn{10}{l}{\ \ \ \ \ \ \ \ \ \ \ \ \
Magnification Ratio: ~${\cal M}_i = (a_{i+1} - a_i)/(a_i - a_{i-1})$.} \\

\end{tabular}
\end{minipage}
\end{center}
\end{table}
%                         END OF TABLE 1
%---------------------------------------------------------------------

On closer inspection of the actual planetary data, we found patterns
other than the Titius--Bode algorithm that could in principle also provide good fits 
to some sections of the data. These patterns are summarized in Table 1 where we 
list the observed semimajor axes $a_i$ of planetary orbits
in our solar system (e.g. Kaufmann 1994) along with three empirical fits to the data, 
the Titius--Bode "law," {\it inversion}, and an arithmetic mean.
The inversion distance for each orbit is the 
geometric mean of the actual semimajor axes of the two nearest neighbors.
The mean distance for each orbit is the arithmetic average of the actual semimajor
axes of the two
nearest neighbors. Relative errors are calculated for all three fits with respect
to the observed values. The inversion distances are listed in Table 1
because this is the first pattern that we saw in the observed data rather than
the Titius--Bode "law." The mean distances are shown for comparison purposes;
they are very accurate only for those orbits in which either or both of the 
other two fits fail. 

The surprisingly small errors in the arithmetic--mean distances of
Venus, Uranus, and Neptune led us to the working hypothesis that the effect
that may cause inversion in the intermediate orbits does not operate at small 
or at large
distances; moreover, it is smoothly replaced at the two ends by a slick new 
regularity that manifests itself as a modest arithmetic average. This smooth 
transition from inversion to arithmetic averaging was another significant
pattern that we saw in the data: For example, the orbit of Uranus 
(along with Jupiter's orbit) reproduces successfully the orbit of Saturn by
inversion while, on the other side, the Saturn--Uranus pair is clearly in arithmetic 
progression with Neptune; and the Earth--Venus pair shows inversion on the
side of Mars and arithmetic progression on the side of Mercury. We thought that
this could not be a numerical coincidence because we found similar smooth
transitions in the orbital distances of the regular (and even some irregular) 
satellites of the Jovian planets.

The geometric--mean spacing of the 1--10 AU objects in Table 1 implies that 
their orbits are {\it inverted images} of every other one with respect 
to the corresponding in--between orbit (e.g., Coxeter 1989), i.e., that the 
semimajor axes of any three consecutive orbits obey the relation
\begin{equation}
a_{i-1} \cdot a_{i+1} \ = \ a_i^2 \ , \ \ \ (3\leq i\leq 7) \ .
\label{d1d3d2d2}
\end{equation}
In geometry, inversion is the gateway to hyperbolic space, where the inverted
orbits would appear to be conveniently equidistant---a perfect symmetry indeed, taking place
in a space that we cannot even visualize. 
But we could not see how the orbital plane of the solar 
system could be so strongly curved in its middle and so flat at the two ends;
so eventually we abandoned this line of reasoning and the
high degree of symmetry in hyperbolic space. 

Next we turned 
to geometric optics, the only part of physics where the same relation occurs. 
Eq.~(\ref{d1d3d2d2}) can also be written as
\begin{equation}
\frac{1}{a_i - a_{i-1}} \ - \ \frac{1}{a_{i+1} - a_i} \ = \ \frac{1}{a_i} \ .
\label{mirror}
\end{equation}
This is a {\it mirror equation} and implies
that the $(i-1)$ and the $(i+1)$ orbits are mirror images of one another 
while the in--between orbit plays the role of a concave mirror. Also,
the magnification ratio
\begin{equation}
{\cal M}_i \ \equiv \ \frac{a_{i+1} - a_i}{a_i - a_{i-1}} \ ,
\label{mag}
\end{equation}
is clearly larger than 1 and approximately constant throughout the system of orbits 
in geometric progression (i.e., for $3\leq i\leq 7$); and it reduces 
to ${\cal M}_i\approx 1$ for those orbits in arithmetic progression (see Table 1). This model 
appeared promising for a while because it suggested that some mechanism could 
potentially be responsible for tapering off the geometric progression to an 
arithmetic progression, as indicated by the magnifications of the orbits.
Unfortunately, the mirror equation of geometric optics is not based on a 
fundamental principle 
per se, it is only a linear approximation valid for paraxial rays, and
we did not think that its deeper underlying principle (Fermat's principle) 
could be applicable to planetary orbits. Because a mirror equation 
similar to eq.~(\ref{mirror}) is not derived from first 
principles in any other part of physical theory, we finally became convinced
that the observed planetary distances, the Titius--Bode algorithm,
and the inversion distances could not be derived from a physical principle;
and we redirected our effort toward mathematical reasons that might be responsible
for the observed order in the solar system. This prompted us to return to the
Lane--Emden equation and to focus exclusively on its intrinsic properties in 
the cases with and without rotation.

In the case of no rotation or for some specific rotation profiles, the 
Lane--Emden equation is scale invariant and can be transformed to an 
autonomous differential equation (see e.g. Bender \& Orzag [1978] and
Visser \& Yunes [2003]) for the theory and the transformations of such 
differential equations). Scale invariance was also exploited
by Graner \& Dubrulle (1994) and Dubrulle \& Graner (1994) who argued that cold, 
self--gravitating, perfect--fluid disks are scale invariant and that this condition
is sufficient to generate unstable radial modes that are equidistant in $\ln R$
(see also Schmitz [1984] and Li et al. [1995] for stability analyses that have 
effectively led to the same result).
Unfortunately, the scale invariance of the inviscid hydrodynamical equations is easily 
broken by the chosen boundary conditions for the equilibrium system
{\it and} for the stability problem; because of this fact, we thought that
we should not look for a geometric progression of the Titius--Bode kind in the
stability problem. After all, scale--invariant unstable modes, such as
those studied by Dubrulle \& Graner (1994), could only produce a generic 
geometric progression and they would be incapable of matching the observed loss 
of inversion at small and large distances, where planetary orbits seem to taper
off neatly to two different arithmetic progressions (see Table 1). 
Therefore, a mechanism
based on such modes of disturbance would suffer from the same problem that also afflicts 
inversion to a hyperbolic space, the Titius--Bode geometric progression, and any other 
idea that overemphasizes the geometric spacing of the intermediate planetary orbits 
and ignores the observed arithmetic progressions of orbits in the inner and the outer 
solar system.

Returning to the scale invariance of the Lane--Emden equation itself, we realized that
it would not matter if this equation lost this symmetry by the applicable
boundary conditions so long as its autonomous form were to exhibit
{\it discrete scale invariance} (DSI; Sornette 1998). In the theory of autonomous 
systems, DSI is a stronger constraint than scale invariance, and it is associated 
with {\it limit cycles} in the phase portrait of the differential
equation (see Appendix B in Visser \& Yunes [2003]). A solution that exhibits DSI
is not invariant for any arbitrary rescaling of the independent variable, but it 
still is self--similar for a specific rescaling of that variable. This fact led 
Sornette (1998) to propose that the Titius--Bode "law" may be a manifestation 
of DSI in the solar system, where the constant ratio of the Titius--Bode geometric
progression or, equivalently, the constant magnification ratio ~${\cal M}_i\approx 2$ ~of 
inversion is the specific factor that rescales each planetary orbit to the next
outward orbit. We followed then the hypothesis of Sornette (1998) 
and the methodology of Visser \& Yunes (2003) and we constructed numerically 
the phase portraits of the autonomous forms of the polytropic and isothermal 
Lane--Emden equations, where we looked for limit cycles and discrete self--similar 
behavior with absolutely no success.

At that point, being convinced that there is no fundamental physics behind the 
observed patterns shown in Table 1 above, and with the theory of nonlinear 
autonomous systems revealing no intrinsic periodicities,
we basically had only one remaining option:
to return to our initial approach and
solve the full boundary--value problem for the differentially--rotating
equilibrium systems in order to see if any interesting patterns would emerge from
the differentially--rotating solutions---patterns similar to the
Bessel--function oscillations found in the uniformly--rotating
$n=1$ polytropic case that did not exist
intrinsically in the differential equation but were generated 
and governed by the applied
boundary conditions. We proceeded to do just that, and our results for the 
isothermal case are described in the sections that follow below. In the end,
we were truly surprised by the simplicity of the mechanism that generates the
patterns seen in Table 1; because the discord between the equilibrium density 
profile favored inherently by the differential rotation and the profile imposed
externally by the boundary conditions, although low--key and inconspicuous, it is 
nonetheless a plain fact and does provide a simple resolution to the long--standing, 
centuries--old problem of planetary "order" in our solar system (see \S~\ref{compo}).

\subsection{Outline}\label{outline}

The remainder of the paper is organized as follows. 
In \S~\ref{models}, we solve analytically for the equilibrium structure 
of the midplane of a gaseous isothermal disk, incorporating in the Lane--Emden 
equation the effects of self-gravity, differential rotation, and thermal pressure.
In \S~\ref{solar}, we adopt a four--parameter analytic solution
as our baseline and we use the rotation profile of the baseline to
compute numerically the corresponding oscillatory equilibrium solution that
obeys physical boundary conditions at the center. Then we fit the density maxima of 
this solution to the planetary orbits in the present solar system in order 
to determine the underlying physical characteristics and the 
stability properties of the baseline model. 
In \S~\ref{discussion}, we discuss the significance of our results for our solar 
system and for protoplanetary disks in general.
%Finally, we summarize our conclusions in \S~\ref{conclusions}.

\section{Isothermal Equilibrium Models}\label{models}

We consider the axisymmetric equilibria that are available to a
rotating self--gravitating gas in the absence of viscosity and
magnetic fields. We adopt cylindrical coordinates ($R, \phi, z$) and
the assumption of cylindrical symmetry ($\partial /\partial z = 0$)
which lets us ignore $z$--dependent gradients and reduces the problem
to one dimension, the distance $R$ from the rotation axis. This
technique has become common practice in studies of rotating,
self--gravitating, fluid disks (e.g., Goodman \& Narayan 1988;
Christodoulou \& Narayan 1992; Christodoulou 1993; Christodoulou,
Contopoulos, \& Kazanas 1996) because it simplifies the stability
analyses of effectively two--dimensional modes of disturbance. In what
follows, we are interested in equilibrium structures
that describe the physical
conditions across the midplane of a protoplanetary disk, so the
assumptions $\partial /\partial\phi = \partial /\partial z = 0$ allow
us to tackle the problem by solving ordinary differential equations (ODEs).

We further adopt a rotation law of the form
\begin{equation}
\Omega (R) = \Omega_0 f(x)\, ,
\label{rot}
\end{equation}
where $x\equiv R/R_0$ is a dimensionless radius and the length scale $R_0$ 
will be specified in eq.~(\ref{length}) below. Furthermore, $\Omega (R)$ is the angular 
velocity, $\Omega_0$ is the value of $\Omega$ at some fixed radius, and the
dimensionless function $f(x)$ for differential rotation is generally an
arbitrary function of $x$. For centrally condensed models, it is
convenient to choose $\Omega_0 = \Omega (0)$ and the regularity condition
$f(0)=1$, while for singular or annular models, we choose 
$\Omega_0 = \Omega (R_0)$ and the normalization $f(1)=1$.

Finally, we assume an isothermal equation of state of the form
\begin{equation}
P = c_0^2 \rho\, ,
\label{state}
\end{equation}
where $P$ is the thermal pressure, $\rho$ is the gas density, 
and $c_0$ is the constant isothermal sound speed.
Additional calculations in which we used polytropes with indices
$n = 1-3$ and the results of Schneider \& Schmitz (1995) who used
polytropes with indices $n < -1$
demonstrate that the characteristics of the solutions presented below
are largely insensitive to the particular choice of $n$.

\subsection{The Lane--Emden Equation With Rotation}\label{LER}

Axisymmetric and cylindrically symmetric, nonmagnetic equilibria for a perfect fluid 
are described by the equation
\begin{equation}
\frac{1}{\rho}\frac{dP}{dR} + \frac{d\Phi}{dR} = \Omega^2 R\, ,
\label{hydro}
\end{equation}
where the gravitational potential $\Phi (R)$ satisfies Poisson's equation
\begin{equation}
\frac{1}{R}\frac{d}{dR}R\frac{d\Phi}{dR} = 4\pi G\rho\, ,
\label{poisson}
\end{equation}
where $G$ is the gravitational constant. Combining
eqs.~(\ref{rot})--(\ref{poisson}) and using the definition $x\equiv R/R_0$, 
we find a second--order nonlinear innonhomogeneous ODE that can be cast in 
the form
\begin{equation}
\frac{1}{x} \frac{d}{dx} x \frac{d}{dx}\ln\tau \ + \ \tau \ = \ 
\frac{\beta_0^2}{2x}\frac{d}{dx}\left(x^2 f^2\right)\, ,
\label{main1}
\end{equation}
where ~$\tau\equiv\rho /\rho_0$, ~$\rho_0$ is the maximum 
density or a fixed cutoff density for singular models,
~$\beta_0\equiv \Omega_0 /\Omega_J$, ~$\Omega_J^2\equiv 2\pi G\rho_0$, ~and
\begin{equation}
R_0^2 \equiv \frac{c_0^2}{4\pi G\rho_0} = \frac{c_0^2}{2\Omega_J^2}\, .
\label{length}
\end{equation}
The term $\Omega_J$ represents the gravitational (Jeans) frequency and 
the dimensionless rotation parameter $\beta_0$ measures centrifugal
support against self--gravity; in general, $0\leq\beta_0\leq 1$, since the 
gas is also partially supported by pressure gradients in the radial direction.

Eq.~(\ref{main1}) reduces to the classical isothermal Lane--Emden equation 
in the absence of rotation ($\beta_0 = 0$).\footnote{Or for a flat rotation
curve with $f(x)=1/x$. This singular rotation law is not of interest in this work.} 
We derive analytically the nonrotating solutions
in \S~\ref{norot} using the modern theory of nonlinear ODEs (e.g., Bender \& Orzag [1978]; 
see also the classic works of Stod\'olkiewicz [1963] and Ostriker [1964]).
Then, in \S~\ref{withrot}, we derive analytically
a class of particular solutions of the full 
problem (eq.~[\ref{main1}] with arbitrary $f(x)$ differential rotation)
and, in \S~\ref{compo}, we discuss a subset of composite power--law solutions 
that are astrophysically interesting despite the fact that they are incapable of
obeying the proper boundary conditions at $x=0$.

\subsection{Nonrotating Solutions}\label{norot}

In the absence of rotation, the isothermal Lane-Emden equation
(eq.~[\ref{main1}] with $\beta_0 = 0$) reads
\begin{equation}
\frac{1}{x} \frac{d}{dx} x \frac{d}{dx}\ln\tau \ + \ \tau \ = \ 0 \, .
\label{main2}
\end{equation}
This equation is scale invariant under the transformation 
($x\to \lambda~x$, ~$\tau\to \lambda^{\textstyle p}~\tau$), where $\lambda$ is an 
arbitrary 
constant and $p=-2$. Therefore, it can be transformed to an autonomous form:
Using this value of $p$, we define $\tau = x^{-2}~w(x)$ 
and we cast eq.~(\ref{main2}) into an equidimensional--in--$x$ equation for the
new function $w(x)$. This equation is invariant under the transformation 
$x\to \lambda~x$ (i.e., its $p=0$) and reads
\begin{equation}
x^2\frac{d^2}{dx^2}\ln w \ + \ x\frac{d}{dx}\ln w \ + \ w \ = \ 0 \, .
\label{main3a}
\end{equation}
Finally, we let $y(x) = \ln w$ and we Euler--transform the independent
variable ($x = e^{\textstyle t}$) to obtain the autonomous form for the function
$y(t)$:
\begin{equation}
\ddot{y} \ + \ e^{\textstyle y} \ = \ 0 \, ,
\label{main4}
\end{equation}
where the dots denote derivatives with respect to $t$.
Since the first derivative is missing from eq.~(\ref{main4}), we can integrate
twice. The first integral is
\begin{equation}
\dot{y}^2 \ = \ C_1 \ - \ 2e^{\textstyle y} \, ,
\label{int1}
\end{equation}
and the solution is
\begin{equation}
\int{\frac{dy}{\sqrt{C_1 - 2e^{\textstyle y}}}} \ = \ C_2 \ \pm \ t \, ,
\label{int2}
\end{equation}
where $C_1$ and $C_2$ are integration constants. The integral
in eq.~(\ref{int2}) can be calculated in closed form
\begin{equation}
\int{\frac{dy}{\sqrt{C_1 - 2e^{\textstyle y}}}} \ = \ \frac{1}{\sqrt{C_1}}
\ln\frac{\sqrt{1 - 2e^{\textstyle y}/C_1}-1}{\sqrt{1 - 2e^{\textstyle y}/C_1}+1} \, ,
\label{int3}
\end{equation}
and a series of backsubstitutions produces the following 
general solution for the density $\tau (x)$:
\begin{equation}
\tau (x) \ = \ 2 A k^2 \frac{x^{k-2}}{(1 + A x^k)^2} \, ,
\label{sol1}
\end{equation}
where $A$ and $k$ are arbitrary {\it positive} constants. The condition
that $A > 0$ is physical and ensures that the density profiles are 
nonnegative. The condition that $k > 0$ is not a limitation: eq.~(\ref{sol1})
is invariant under the transformation ($k\to -k$, ~$A\to 1/A$) because $k$
contains implicitly the $\pm$ duality seen in eq.~(\ref{int2}) above.
Therefore, only positive values of $k$ need to be considered without this
causing loss of generality.

The equilibrium solutions obtained in eq.~(\ref{sol1}) can be classified 
into three types:

\begin{enumerate}
\item For $0<k<2$, the density profiles are singular at $x=0$ and decrease 
monotonically for $x>0$. These solutions are analogues of the singular isothermal
sphere. 
\item For $k=2$, the density profile is centrally condensed and can easily
satisfy proper boundary conditions at $x=0$. Using $A=1/8$, eq.~(\ref{sol1})
reduces to the Stod\'olkiewicz--Ostriker solution
\begin{equation}
\tau (x) \ = \ \frac{1}{(1 + x^2/8)^2} \ ,
\label{SO}
\end{equation}
for which $\tau (0) = 1$ and 
$\frac{d\tau}{dx} (0) = 0$.
\item For $k>2$, each density profile has a hole at $x=0$---i.e., 
$\tau (0)=0$---and peaks at a finite radius $x_*$, namely
\begin{equation}
x_* \ = \ \left(\frac{1}{A}\cdot\frac{k-2}{k+2}\right)^{1/k} \ ,
\label{xcr}
\end{equation}
where $\tau (x_*) = 1$ and $\frac{d\tau}{dx} (x_*) = 0$; this normalization fixes 
the value of $A$ for any choice of $k>2$. The two constants are related 
by the equation
\begin{equation}
A^2 \ (k-2)^{k-2} \ (k+2)^{k+2} \ = \ 2^k \ . 
\label{Ak}
\end{equation}
The mass in all of these models is strongly concentrated around the density maximum
$x=x_*$; this gives them the appearance of slender annuli despite the presence
of extended regions of very low densities on either side of the maximum; regions
that extend all the way to $x=0$ and to $x=\infty$. Also note that the solutions 
for $k>3$ are concave and shallow near the center where they have, not only $\tau (0)=0$, 
but also $\frac{d\tau}{dx} (0) = 0$; while the solutions for $2<k\leq 3$ ~are 
convex and steeply rising near the center where they have ~$d\tau /dx \to 2Ak^2(k-2)/x^{3-k}$
~as ~$x\to 0$. 
\end{enumerate}

\noindent
At large radii ($x>>1$), all solutions are decreasing with radius 
and the density falls off as $x^{-k-2}$. This rapid asymptotic decline,
which is steeper than $x^{-2}$,
is responsible for keeping the mass per unit length $\mu$ finite in all models.
A direct integration of eq.~(\ref{sol1}) shows that 
$\mu /2\pi\rho_0 R_0^2 = 2k$, independent of the constant $A$.
Letting $k=2$ in this equation, we recover Ostriker's (1964) result,
$\mu = 8\pi\rho_0 R_0^2$, for the centrally condensed cylinder.

\subsection{Rotating Solutions}\label{withrot}

The isothermal Lane-Emden equation with differential rotation
(eq.~[\ref{main1}]) is repeated here for the purpose of discussion:
\begin{equation}
\frac{1}{x} \frac{d}{dx} x \frac{d}{dx}\ln\tau \ + \ \tau \ = \ 
\frac{\beta_0^2}{2x}\frac{d}{dx}\left(x^2 f^2\right)\, .
\label{main3}
\end{equation}
When the right--hand side (hereafter RHS) of this equation is nonzero
(i.e., when $\beta_0\neq 0$ and $f(x)\neq 1/x$), the property
of scale invariance is lost from all cases of interest (uniform rotation,
power--law rotation, etc.), irrespective of the prescription chosen for the 
differential rotation function $f(x)$.\footnote{Eq.~(\ref{main3}) with a nonzero
RHS is scale invariant only for $f(x) = \sqrt{A\ln x + B}/x$, where $A$ and $B$
are arbitrary constants. This case can be solved by transforming the scale--invariant
ODE to its autonomous form (as was explained in \S~\ref{norot}), but there is no need
to do so; its solution is obtained easier by the method described in this
subsection.} 
Eq.~(\ref{main3}) has no special symmetry associated with it, 
and this is enough to make many researchers turn the other way. This is probably why 
some interesting features of eq.~(\ref{main3}) that we describe below have gone 
unnoticed for so long.

The RHS of eq.~(\ref{main3}) is not merely a rotation--dependent correction term to the 
classical isothermal Lane--Emden equation~(\ref{main2}). The introduction of rotation 
changes the properties of the ODE to such a large extent that the nonrotating solutions 
found in \S~\ref{norot} cannot guide the effort to find rotating equilibrium 
solutions.
In fact, it is the functional form of the RHS that determines now the structure of the 
solutions of the entire ODE: By inspection of eq.~(\ref{main3}), we can write down 
an entire class of particular equilibrium solutions, namely
\begin{equation}
\tau (x) \ = \ \frac{\beta_0^2}{2x}\frac{d}{dx}\left(x^2 f^2\right)\, ,
\label{part1}
\end{equation}
provided that
\begin{equation}
\frac{d}{dx} x \frac{d}{dx}\ln\tau \ \equiv \ 0 \ ,
\label{part2}
\end{equation}
also holds true. Using eq.~(\ref{part1}), we write
\begin{equation}
\ln\tau \ = \ \ln\frac{\beta_0^2}{2} \ - \ \ln x \ + \ 
\ln\frac{d}{dx}\left(x^2 f^2\right)\, ,
\label{part3}
\end{equation}
and substituting this form into eq.~(\ref{part2}) we find an ODE
for all the differential--rotation laws $f(x)$ that satisfy eq.~(\ref{part2}) 
identically and make eq.~(\ref{part1}) a family of exact solutions of
the Lane--Emden equation with rotation (eq.~[\ref{main3}]):
\begin{equation}
\frac{d}{dx} x \frac{d}{dx} \ln \frac{d}{dx}\left(x^2 f^2\right) \ = \ 0 \, .
\label{part4}
\end{equation}
This third--order linear ODE can be integrated directly to yield 
the following results:
\begin{equation}
\frac{d}{dx}\left(x^2 f^2\right) \ = \ A x^k \ ,
\label{class1}
\end{equation}
implying that
\begin{equation}
\tau (x) \ = \ \frac{\beta_0^2}{2}\cdot A x^{k-1} \ ,
\label{class2}
\end{equation}
and
\begin{equation}
f(x) \ = \ \frac{\sqrt{Ag(x) + B}}{x} \ ,
\label{class3}
\end{equation}
where $A$, $B$, and $k$ are arbitrary integration constants (that are unrelated
to those used in \S~\ref{norot}) and
\begin{equation}
g(x) \ \equiv \  \left\{ \begin{array}{cc} 
         x^{k+1}/(k+1) \ , & \ {\rm if} \ \ \ k \neq -1 \\
         \ln x \ , \ \ \ \ \ \ \ \ \ \ \ & \ {\rm if} \ \ \ k = -1 
         \end{array} \right. \ ,
\label{class4}
\end{equation}
implying that ~$dg/dx = x^k$ ~for all values of $k$.
With so many free parameters ($A$, $B$, and $k$) in the differential--rotation 
profile, these results can easily become a theorist's playground. Here we highlight 
just a few interesting points:

\begin{enumerate}
\item {\it Parameter Constraints}.---Eq.~(\ref{class2}) shows that 
$\tau (x) > 0$ only for $A > 0$. This constraint also limits the physical
values of $k$ when $B \leq 0$ in eq.~(\ref{class3}); for example, 
$k\geq -1$ when $B=0$. This limitation can be easily circumvented by 
implementing composite rotation profiles with $B>0$ (see item 4 in this list 
and \S~\ref{compo} below).

\item {\it Monotonically Decreasing Profiles}.---Eq.~(\ref{class2}) shows that
$\tau (x)$ is a decreasing function of $x$ for $|k| < 1$. The same condition
is sufficient to also make $f(x)$ a decreasing function of $x$ provided that
$B\geq 0$ in eq.~(\ref{class3}).

\item {\it Uniform Rotation}.---For $A=2$, $B=0$, and $k=1$, eq.~(\ref{class3})
reduces to $f(x)=1$ and the equilibrium density (eq.~[\ref{class2}]) then is
~$\tau (x) = \beta_0^2 = {\rm constant}$. Note that this constant cannot be
adjusted freely, e.g., it cannot be reset to 1; 
the requirement that $f(x)=1$ fixes $A$ in eq.~(\ref{class3})
and then the uniform density gets fixed to $\beta_0^2$ by eq.~(\ref{class2}).  

\item {\it Composite Profiles With} $B>0$.---Steep density profiles 
with $k < -1$ can be obtained by selecting $B > 0$ and by incorporating a 
central region of uniform rotation (see \S~\ref{compo} for details). 
Even more complex equilibrium profiles
can be constructed by combining two or more power laws. A composite
profile is demonstrated in \S~\ref{solar1} below, where we connect two
disjoint regions of constant density with a power law and we apply the result
to the early structure of the solar nebula.

\item {\it Asymptotic Regime}.---For $k < -1$ and $B>0$, eq.(\ref{class4})
shows that $g(x)\to 0$ as $x\to\infty$ and eq.(\ref{class3}) then exhibits  the
asymptotic behavior $f(x)\to \sqrt{B}/x$. Therefore, all steep density profiles
with $k < -1$ and $\tau (x)\propto x^{k-1}$ approach a flat rotation curve 
($\Omega R\to$ constant) at large radii, independent of the value of $k$.
\end{enumerate}

From the perspective of the physics that dictates the above profiles, 
the solutions~(\ref{part1}) of the Lane--Emden equation~(\ref{main3}) describe a 
class of rotating self--gravitating equilibria in which $z$--gradients are neglected
and the {\it radial gradient} of the 
gravitational acceleration is balanced exactly by the {\it radial gradient} 
of the centrifugal acceleration at every radius $R$. This occurs because, in the Lane--Emden
equation, we have gone to second order by taking an extra derivative on the components of the 
equation of hydrostatic equilibrium. The balance of gradients can be seen, most easily, by 
substituting eq.~(\ref{part1}) into the one--dimensional Poisson's equation ${\bf\nabla}^2\psi = \tau$, 
where $\psi\equiv\Phi / c_0^2$ is the normalized potential; the result is
\begin{equation}
\frac{1}{x}\frac{d}{dx} x \left[\frac{d\psi}{dx}\right]
\ = \ \
\frac{1}{x}\frac{d}{dx} x \left[\frac{1}{2}\beta_0^2 \cdot x f^2\right] 
\, .
\label{gradients}
\end{equation}
In this equation, the bracketed terms are the gravitational and centrifugal accelerations, 
respectively. This type of balance is different than the balance commonly discussed between the
magnitudes of these two accelerations in rotating gravitating systems;
and the power--law density solutions are borne out of this conformance of 
the two gradients, whereas the familiar stalemate between centrifugal and gravitational 
force is only relevant to purely homogeneous fluid equilibrium systems or particle systems 
with no pressure support. In the isothermal gaseous case of interest here, a
power--law density profile satisfies naturally the condition that the {\it radial variation}
of the enthalpy gradient ~$\rho^{-1}dP /d\ln R$ ~be zero (see eq.~[\ref{part2}]) 
and so the pure power--law profile is not at all influenced by the radial variation of the 
pressure gradient---it is an exact solution of eq.~(\ref{main3}).\footnote{In the 
polytropic case, however, the analogue of eq.~(\ref{part2}), namely
$$
\frac{d}{dR}\left(\rho^{-1}\frac{dP}{d\ln R}\right) \ = \ 0\ ,
$$
is not satisfied by pure power--law profiles 
(except in the trivial case with ~$P = {\rm constant}$); then, 
pressure--gradient variations do affect the structure of the underlying equilibrium 
solutions, but not in a dramatic fashion.}

\subsection{Composite Models and Boundary Conditions}\label{compo}

Many of the rotation profiles discussed above are singular at $x=0$. The solutions 
for the density, especially, are all pure power laws and, for $k<1$, 
they all diverge as $x\to 0$.
As was mentioned in \S~\ref{withrot}, this is not a serious problem because the
singularity at the center can be removed by assuming that the central region rotates
uniformly and that the density profile switches to a power law beyond a "core" radius
$x=x_1$. The core radius $x_1$ can be chosen freely even for steep density profiles
with $k < -1$, but then the constant $B$ in eq.~(\ref{class3}) must be positive and 
it should be adjusted accordingly so that $f(x)$ is everywhere positive and 
monotonically decreasing with $x$. It turns out that the physical requirement
that $f(x)\geq 0$ is weaker than the monotonicity condition that $df/dx\leq 0$.
With $A>0$ to ensure that $\tau (x) > 0$ and assuming that $B > 0$ and $k < -1$, 
we find from eq.~(\ref{class3}) that ~$f(x)\geq 0$ ~if
\begin{equation}
B \ \geq \ \frac{A}{\ell x_1^{\textstyle\ell}} \ ,
\label{cond1}
\end{equation}
where ~$\ell\equiv |k+1| > 0$; ~and the stronger condition that ~$df/dx\leq 0$ ~if
\begin{equation}
B \ \geq \ \frac{A}{\ell x_1^{\textstyle\ell}}\left(1 + \frac{\ell}{2}\right) \ .
\label{cond2}
\end{equation}
Therefore, all composite models with a uniformly--rotating core region $x\leq x_1$ 
must satisfy the stronger condition~(\ref{cond2}) for any choice of $A>0$ and 
$k < -1$ in their equilibrium density profiles (eq.~[\ref{class2}]).

The uniformly--rotating cores of the composite models discussed above call attention to
another interesting feature: The density in these models must be constant 
and equal to $\beta_0^2$ in order to support this type of rotation (see also
item 3 in the list of \S~\ref{withrot}).
So these models cannot obey the boundary condition that ~$\tau (0) = 1$
~for centrally condensed structures. More generally, all the power--law solutions
that we derived for the density in \S~\ref{withrot} and all the composite models, 
although they are exact intrinsic solutions of the Lane--Emden equation with rotation 
(eq.~[\ref{main3}]), they do {\it not} solve the associated boundary--value
problem.
The question then is: How are the actual density profiles of centrally condensed
equilibrium models going to behave once the proper set of boundary conditions 
\{$\tau (0)=1$, ~$\frac{d\tau}{dx} (0)=0$\} are imposed at the center? 

The answer
can be obtained by numerical integrations that enforce the desired central 
boundary conditions in eq.~(\ref{main3}) and in the analogous polytropic 
Lane--Emden equation with rotation; and by examining the analytic solution
to the full boundary--value problem of the linear polytropic case with index $n=1$ 
and uniform rotation. All the numerical (isothermal and polytropic) solutions, 
as well as the $n=1$ analytic solution,\footnote{For uniformly--rotating
polytropic cylinders with $n=1$, the solution to the boundary--value problem 
\{$\tau (0)=1$, ~$\frac{d\tau}{dx} (0)=0$\} is analytic (Robe 1968):
$$
\tau (x) \ = \ \left(1 - \beta_0^2\right) J_0 (x) \ + \ \beta_0^2\, ,
$$
where $J_0 (x)$ is the zeroth--order Bessel function of the first kind
and all the other symbols are defined as in this work. (However, for
$n=1$ polytropes, $c_0^2\equiv 2K\rho_0$, where $K=P/\rho^2$ is the 
polytropic constant, and $R_0^2\equiv K/2\pi G$.) It is clear in this 
solution that the Bessel function oscillates permanently about the 
$\tau = \beta_0^2$ line, which is a particular solution of the $n=1$ 
linear ODE.}
demonstrate routinely (see also Schneider \& Schmitz [1995])
that the equilibrium density profiles lose their
monotonicity the first time they cross below the corresponding 
particular solutions (solutions analogous to those discussed in 
\S~\ref{withrot} for the isothermal models). Once the first such 
crossing occurs at some relatively small radius (see Figs.~\ref{fig2} and~\ref{fig3} below), 
the actual physical solutions recognize the existence of
the corresponding intrinsic solutions and they turn and oscillate
permanently about the density level defined by these particular solutions.
This of course happens because the particular solutions are fundamental
"baseline" solutions of the ODE itself, regardless of externally--imposed
conditions. When a set of external conditions are imposed at $x=0$ for 
physical reasons, the actual solutions emerging from the center do not match 
the baseline solutions (since the power--law behavior of the baseline 
is incompatible with the imposed conditions), but they are 
nonetheless attracted to them because the baseline solutions satisfy 
the ODE inherently. The result then is a permanent mismatch
around the baseline that extends over all radii.
This behavior is demonstrated in \S~\ref{solar2} below,
in the example model shown in Fig.~\ref{fig2} and, notably, in the composite 
isothermal model of the solar nebula  shown in Fig.~\ref{fig3}.

\section{Application to Our Solar System}\label{solar}

The oscillatory behavior of the density profiles discussed in \S~\ref{compo} 
finds a natural application to the structure of the midplane of the 
solar nebula. For this application, we need a composite model as a baseline 
because such models allow for equilibrium
density power laws that can be arbitrarily steep. To this end, we formulate 
in \S~\ref{solar1} a composite model using the exact analytic solutions determined 
in \S~\ref{withrot} above. Then, in \S~\ref{solar2}, we solve numerically the
corresponding boundary--value problem that exhibits the same rotation profile as
the analytic model; and we obtain an oscillatory solution for the density profile
subject to the applied physical boundary conditions.
Finally, we proceed to fit the density peaks of this model to the observed planetary 
distances in our solar system and we conclude in \S~\ref{solar3} by determining
important physical parameters associated with the structure, the dynamics, and 
the stability of the solar nebula.

\subsection{Composite Equilibrium Model}\label{solar1}

For our baseline equilibrium model of the midplane of the solar nebula, 
we adopt a composite analytic solution in which the isothermal density profile 
has the form of a truncated power law:
\begin{equation}
\tau_{_{\textstyle base}} (x) \ = \ \beta_0^{\textstyle 2} \ \cdot 
         \left\{ \begin{array}{cc} 
         1 \ , & \ {\rm if} \ \ \ x \leq x_1 \\
         (x_1/x)^{\textstyle\delta} \ , & \ \ \ \ \ \ \ \, {\rm if} \ \ \ x_1 < x < x_2 \\
         (x_1/x_2)^{\textstyle\delta} \ , & \ {\rm if} \ \ \ x \geq x_2
         \end{array} \right. \ ,
\label{den3}
\end{equation}
where $x_1$ is the radius of the constant--density core region, 
$x_2$ is the truncation radius beyond which the density remains constant at a low value, 
and the power--law index $\delta$ is defined by  
\begin{equation}
\delta \ \equiv \ 1 - k \ \neq \ 2 \ .
\label{ddd}
\end{equation}
The condition that $\delta\neq 2$ implies that $k\neq -1$ and excludes the logarithmic 
rotation laws from consideration: for $k\neq -1$, logarithms 
are not introduced in the general form of the rotation law (eq.~[\ref{class3}]) by the 
lower branch of eq.~(\ref{class4}). Moreover, we consider below  an even more limited 
range of indices, namely ~$\delta > 2$, ~since we are primarily interested in steep 
density profiles with ~$k < -1$. 

Introducing $x_1$ and $x_2$ in the above profile is equivalent to
specifying three different values for the constant $A$ in eq.~(\ref{class2}) of 
\S~\ref{withrot}. These values are chosen so that the density profile remains 
continuous as it switches from one branch to the next, namely
\begin{equation}
A \ = \ 2 \ \cdot \left\{ \begin{array}{cc} 
         1 \ , & \ {\rm if} \ \ \ x \leq x_1 \\
         x_1^{\textstyle\delta} \ , & \ \ \ \ \ \ \ \ {\rm if} \ \ \ x_1 < x < x_2 \\
         (x_1/x_2)^{\textstyle\delta} \ , & \ {\rm if} \ \ \ x \geq x_2
         \end{array} \right. \ .
\label{den30}
\end{equation}
The rotation law can then be determined from eqs.~(\ref{class3}) and~(\ref{den30}) 
by finding the values of the constant $B$ that also make this profile
continuous at the junctions where $x=x_1$ and $x=x_2$, namely
\begin{equation}
B \ = \  \frac{\delta}{\delta - 2} \ \cdot \left\{ \begin{array}{cc} 
         0 \ , & \ {\rm if} \ \ \ x \leq x_1 \\
         x_1^{\textstyle 2} \ , 
         & \ \ \ \ \ \ \ \, {\rm if} \ \ \ x_1 < x < x_2 \\ 
         \left[\left(x_1/x_2\right)^{\textstyle 2}
                        - \left(x_1/x_2\right)^{\textstyle\delta}\right] 
         x_2^{\textstyle 2} \ , 
         & \ {\rm if} \ \ \ x \geq x_2
         \end{array} \right. \ .
\label{den300}
\end{equation}
As ~$x\to x_1$ ~from the right, the value of $B$ in the intermediate branch satisfies 
marginally the monotonicity requirement (eq.~[\ref{cond2}]) determined in \S~\ref{compo} 
above.

In practice,
it is easier to integrate the differential equation~(\ref{class1}) in
each of the three regions of the model and use the integration constants
along with eq.~(\ref{den30}) to ensure continuity at the junctions $x_1$ and $x_2$.
Using either method, we find that the rotation law has the form
\begin{equation}
f (x) \ = \ \left\{ \begin{array}{cc} 
           \ 1 \ , \ \ \ \ \ \ \ \ \ \ \ \ \ \ \ \ \ \
\ \ \ \ \ \ \ \ \ \ \ \ \ \ \ \ \ \ \ \ \ \ \ \ \ \ \ \
\ \ \ \ \ \ \ \ \ \ \ \ \ \ \ \ \ \ \ \ \ \ \ \ \ \
         {\rm if} \ \ \ x \leq x_1 \\
               \\
         \sqrt{\frac{\textstyle 1}{\textstyle\delta - 2}\left[ 
               {\textstyle\delta}\left(x_1/x\right)^{\textstyle 2}
                        - {\textstyle 2}\left(x_1/x\right)^{\textstyle\delta}
                             \right]} \ , 
\ \ \ \ \ \ \ \ \ \ \ \ \ \ \ \ \
  \ \ \ \ \ \ \ \ \ {\rm if} \ \ \ x_1 < x < x_2 \\
                \\ 
         \sqrt{\frac{\textstyle\delta}{\textstyle\delta - 2}
         \left[ \left(x_1/x_2\right)^{\textstyle 2}
                        - \left(x_1/x_2\right)^{\textstyle\delta} 
         \right] \left(x_2/x\right)^{\textstyle 2} +
         \left(x_1/x_2\right)^{\textstyle\delta}}   \ , 
 \ \ \ \ \ \ \ \ {\rm if} \ \ \ x \geq x_2 &
         \end{array} \right. .
\label{den31}
\end{equation}
It is easy to show that this rotation law obeys the physical requirements that
$f(x)>0$ and $df/dx\leq 0$ at all radii for any choice of the parameter
set ~\{$x_1 > 0$, ~$x_2 > x_1$, ~$\delta > 2$\}. ~Notice that, outside the core,
the rotation profile is monotonically decreasing everywhere; and that it becomes
asymptotically flat at very large radii: as ~$x\to\infty$, ~then 
\begin{equation}
f(x)\to \left(\frac{x_1}{x_2}\right)^{\textstyle\delta /2} \, . 
\label{fasym}
\end{equation}
Thus, in contrast to the pure power--law
density profiles (item 5 in \S~\ref{withrot}), this composite profile exhibits 
nearly uniform rotation at very large distances.
This is because the density of the model is not allowed to decrease at large distances; 
instead, it is kept constant at the low level shown in eq.~(\ref{den3}) for $x\geq x_2$.

The differential--rotation function ~$f(x)$ ~of eq.~(\ref{den31})
and the corresponding equilibrium density profile 
~$\tau_{base} (x)/\beta_0^2$ ~of eq.~(\ref{den3}) are shown in Figure~1 
for ~$x_1=100$, ~$x_2=500$, ~and for various choices of the power--law 
index ~$\delta > 2$.

\subsection{Solutions of the Boundary--Value Problem \\
            and Parameter Optimization to Planetary Distances}\label{solar2}

The above composite equilibrium model is characterized by four free parameters: 
the core radius $x_1$, the truncation radius $x_2$, the rotation parameter $\beta_0\leq 1$, 
and the power--law index $\delta > 2$. The density profile (eq.~[\ref{den3}])
of this baseline solution of the Lane--Emden ODE (eq.~[\ref{main3}]) is not capable 
of satisfying physical boundary conditions at the center and it serves only as a mean 
approximation to the density of the corresponding physical model. The general form
of the rotation law of the baseline (eq.~[\ref{den31}]) can, however, be adopted 
for the differential rotation $f(x)$ of the physical model as well. Then 
eq.~(\ref{den3}) provides a prescription for the RHS of the Lane--Emden 
equation~(\ref{main3}) which can thus be written as
\begin{equation}
\frac{1}{x} \frac{d}{dx} x \frac{d}{dx}\ln\tau \ + \ \tau \ = \ 
\tau_{_{\textstyle base}} (x) \, .
\label{main5}
\end{equation}
This ODE can be integrated numerically subject to the physical boundary conditions 
that
\begin{equation}
\left\{ \begin{array}{c} 
         ~~\tau (0) \ = \ 1 \\
         \\
         \frac{\textstyle d\tau}{\textstyle dx} (0) \ = \ 0
         \end{array} \right\} \ .
\label{bc}
\end{equation}
An example is shown in Figure~2 ~for an equilibrium model with ~$x_1=200$, ~$x_2=1000$, 
~$\beta_0 = 0.2$, ~and ~$\delta = 3$. Notice that, in the linear scale of Fig.~2a, the actual 
density peaks (along the solid line) are approximately equidistant in regions where the 
baseline density (dashed line) is uniform; while in the intermediate region, 
they spread farther apart as they are trying to keep up with the steeply declining baseline.
In the logarithmic scale of Fig.~2b, the same peaks appear to come closer together in the two 
areas where the baseline is flat and become equidistant in the middle area along the gradient 
of the baseline.

A far more interesting application is shown in Figure~3. 
The Lane--Emden equation~(\ref{main5}), subject to the physical boundary 
conditions~(\ref{bc}), has been integrated numerically for various choices of the four 
free parameters. The resulting equilibrium profiles have been optimized for the 
present solar system assuming that their density maxima correspond to the 
observed semimajor axes of the planetary orbits out to and including Pluto.
Not counting the central peak at $x=0$, the third 
density maximum is always scaled to a 
distance of 1~AU during this {\it nonlinear unconstrained optimization}. In the best--fit 
model shown in Figure~\ref{fig3}, the third density peak occurs at $x=44.564$, implying that 
the length scale of the solar nebula in its isothermal phase was quite small 
($R_0 = 0.022440$~AU; see also \S~\ref{solar3} below). 

%---------------------------------------------------------------------
%                             TABLE 2
\begin{table}[t]
\begin{center}
\begin{minipage}{7in}
\caption{}
\begin{tabular}{ccccc}
\multicolumn{5}{c}{\sc Locations of Density Peaks} \\
\multicolumn{5}{c}{\sc in the Best--Fit Model} \\
\multicolumn{5}{c}{\sc of the solar Nebula} \\
\hline\hline
Index & Planet  & Semimajor     & Peak       &       \\
 ~$i$ & Name    & Axis          & Location   & Error \\
      &         & $a_i$ (AU)    & $d_i$ (AU) & (\%)  \\
\hline\hline
1     & Mercury & 0.387         & 0.362      & $-6.5$    \\
2     & Venus   & 0.723         & 0.705      & $-2.5$    \\
3     & Earth   & 1             & 1          & . . .     \\
4     & Mars    & 1.524         & 1.588      & ~~4.2     \\
5     & Ceres   & 2.765         & 2.686      & $-2.9$    \\
6     & Jupiter & 5.203         & 4.930      & $-5.2$    \\
7     & Saturn  & 9.537         & 9.843      & ~~3.2     \\
8     & Uranus  & 19.19         & 20.09      & ~~4.7     \\
9     & Neptune & 30.07         & 29.66      & $-1.4$    \\
10    & Pluto   & 39.48         & 39.19      & $-0.7$    \\
11    &         & $^{a}$        & 48.73      &           \\ 
12    &         & $^{b}$        & 58.27      &           \\
13    & Eris    & 67.89 $^{c}$  & 67.80      & $-0.1$    \\ 

\hline\hline
\multicolumn{5}{l}{{\sc Notes}: }\\
\multicolumn{5}{l}{$(a)$~(136472) 2005 FY$_9$ ~is at ~$a=45.66$~AU
($d_{11}$ deviates by 6.7\%). }\\
\multicolumn{5}{l}{$(b)$~(84522) 2002 TC$_{302}$ ~is at ~$a=55.02$~AU
($d_{12}$ deviates by 5.9\%). } \\
\multicolumn{5}{l}{$(c)$~From Brown, Trujillo, \& Rabinowitz (2005). } \\

\end{tabular}
\end{minipage}
\end{center}
\end{table}
%                         END OF TABLE 2
%---------------------------------------------------------------------

The density maxima $d_i$ of the best--fit model have been converted to AU and are listed 
in Table~2 along with the observed orbital semimajor axes $a_i$ taken from Table~1 above. 
The new dwarf planet Eris was not used in the optimization, but it is also listed at the 
bottom of Table~2 for comparison purposes. 
In addition, Table~3 displays all the extrema in the best--fitted oscillatory density 
profile out to 106~AU. Table~3 may be useful to observers who are trying to locate more 
dwarf planets in the outer solar system and to theorists who intend to build more 
sophisticated models of planitesimal accumulation in the solar nebula.

%---------------------------------------------------------------------
%                             TABLE 3
\begin{table}[t]
\begin{center}
\begin{minipage}{7in}
\caption{}
\begin{tabular}{ccccc}
\multicolumn{5}{c}{\sc Density Extrema} \\
\multicolumn{5}{c}{\sc in the Best--Fit Model} \\
\multicolumn{5}{c}{\sc of the solar Nebula} \\
\hline\hline
Index & Location       & Minimum             & Location       & Maximum          \\
 ~$i$ & of Minimum     & Density             & of Maximum     & Density          \\
      & $d_{min}$ (AU) & $\tau (d_{min})$    & $d_{max}$ (AU) & $\tau (d_{max})$ \\
\hline\hline
  1 & 0.17901  & $7.414\times 10^{-2}$ & 0.36162  & $2.809\times 10^{-1}$ \\ 
  2 & 0.53124  & $1.069\times 10^{-1}$ & 0.70474  & $2.466\times 10^{-1}$ \\ 
  3 & 0.90231  & $1.141\times 10^{-1}$ & 1        & $1.216\times 10^{-1}$ \\ 
  4 & 1.4282   & $3.571\times 10^{-2}$ & 1.5875   & $3.757\times 10^{-2}$ \\ 
  5 & 2.4182   & $9.483\times 10^{-3}$ & 2.6858   & $9.821\times 10^{-3}$ \\ 
  6 & 4.4935   & $2.023\times 10^{-3}$ & 4.9295   & $2.056\times 10^{-3}$ \\ 
  7 & 9.7814   & $3.165\times 10^{-4}$ & 9.8431   & $3.165\times 10^{-4}$ \\ 
  8 & 15.268   & $1.586\times 10^{-4}$ & 20.093   & $2.835\times 10^{-4}$ \\ 
  9 & 24.855   & $1.705\times 10^{-4}$ & 29.661   & $2.714\times 10^{-4}$ \\ 
 10 & 34.421   & $1.773\times 10^{-4}$ & 39.193   & $2.642\times 10^{-4}$ \\ 
 11 & 43.962   & $1.818\times 10^{-4}$ & 48.726   & $2.593\times 10^{-4}$ \\ 
 12 & 53.490   & $1.851\times 10^{-4}$ & 58.266   & $2.557\times 10^{-4}$ \\ 
 13 & 63.028   & $1.876\times 10^{-4}$ & 67.800   & $2.529\times 10^{-4}$ \\ 
 14 & 72.557   & $1.897\times 10^{-4}$ & 77.336   & $2.507\times 10^{-4}$ \\ 
 15 & 82.092   & $1.914\times 10^{-4}$ & 86.857   & $2.488\times 10^{-4}$ \\ 
 16 & 91.626   & $1.928\times 10^{-4}$ & 96.399   & $2.473\times 10^{-4}$ \\ 
 17 & 101.15   & $1.940\times 10^{-4}$ & 105.91   & $2.460\times 10^{-4}$ \\ 
\hline\hline

\end{tabular}
\end{minipage}
\end{center}
\end{table}
%                         END OF TABLE 3
%---------------------------------------------------------------------

Looking at orbital distances interior to the orbit of Eris in Table~2, we see that 
there are two peaks in the model, $d_{11} = 48.73$~AU and $d_{12} = 58.27$~AU, in which 
large dwarf planets have not been discovered. These two orbital distances lie beyond the 
outer "edge" or "gap" of the classical Kuiper belt (47--48~AU; e.g., Delsanti \& Jewitt 2006),
in an area where the number of orbiting objects decreases dramatically (e.g., Morbidelli, 
Brown, \& Levison 2003). Despite that, both peaks are located to within $\sim$6\% from two 
large objects: $d_{11}$ is near Makemake (2005 FY$_9$), the third largest 
classical--Kuiper--belt object after Pluto and Haumea; and $d_{12}$ is 
near 2002 TC$_{302}$, the second largest scattered--disk object after Eris.

The relative errors for each individual density peak of the model are also listed in 
Table~2. We see that the largest relative error is ~$-6.5$\% ~for the first peak that 
corresponds to the orbit of Mercury. This deviation of $\sim$0.025~AU is still quite 
small by solar--system standards.
The optimization algorithm has minimized the mean relative error $\sigma$ for the 
first 10 orbits listed in Table~2. This was defined as a "standard deviation"
by using the square 
deviations of 9 density peaks and excludes the peak that corresponds
to the Earth's orbit which has zero deviation because of our scaling assumption
that $d_3\equiv 1$~AU.
Thus:
\begin{equation}
\sigma \ \equiv \ \sqrt{\frac{1}{N-2}\cdot\sum_{i=1}^{N}\frac{(d_i - a_i)^2}{a_i^2}} \ ,
\label{error}
\end{equation}
where $N=10$ and the $i=3$ term does not contribute to the sum ($d_3 - a_3\equiv 0$). 
By construction, this definition of the mean relative error
places more weight to the orbital distances of planets near the Sun 
and allows for larger errors in the locations of the outer density peaks.
Despite this skewing of the fit, the mean relative error 
for the best--fit model is $\sigma =$~4.1\%, 
affirming that our simple equilibrium model succeeds in matching all of
the observed planetary distances very well.

\subsection{Physical Parameters of the Solar Nebula}\label{solar3}

The parameters of the best--fit model determined by the optimizing algorithm
along with the scaling assumption that $d_3 = 1$~AU (see Table 2)
constitute a set of important dynamical parameters of the 
long--gone solar nebula:
\begin{equation}
\left\{ \begin{array}{c} 
         \beta_0 \ = \ 0.41465   \\
         \delta \ \ = \ 2.5362  \\
         k \ = \ 1 - \delta \ = \ -1.5362  \\
         $---------------------------$ \\
         R_0 \ = \ 0.022440~~{\rm AU}  \\
         x_1 \ = \ 36.511 \ := \ 0.81931~~{\rm AU}    \\
         x_2 \ = \ 505.45\, \ := \ 11.342~~{\rm AU}  \\
         \end{array} \right\} \ .
\label{neb}
\end{equation}

The value of the rotation parameter $\beta_0\approx 0.4$ indicates that the rotation 
of the isothermal nebula was moderate, about 40\% of the maximum value
allowed by self--gravity. This result also indicates that the best--fit model
(and the other models discussed here) is stable to nonaxisymmetric disturbances
because $\beta_0$ is much below the critical value of $\beta_* = 0.7$. 
This critical value can be obtained from the $\alpha$-parameter 
criterion for stability of rotating, self--gravitating, gaseous systems
($\alpha\leq 0.35$; Christodoulou, Shlosman, \& Tohline 1995) by combining the definition 
$\beta_0\equiv\Omega_0 /\Omega_J$ with the definition of $\alpha$ for disks 
where $\alpha\equiv\Omega_0 /2\Omega_J$ to get the relation 
$\beta_0 = 2 \alpha$ which implies that $\beta_* = 0.7$.
Of course the models discussed in this section are stable to axisymmetric 
disturbances as well, since they all satisfy the Rayleigh criterion.

The value of 
the power--law index $\delta\approx 2.5$ indicates that the density profile of the
differentially--rotating region of the nebula declined with radius, on average, as
$R^{-2.5}$. This region extended from a radius of $x_1\approx 0.8$~AU out to a radius 
of $x_2\approx 11.3$~AU. Of these parameters, only $x_2$ is slightly uncertain because the
mean relative error $\sigma$ (eq.[\ref{error}]) places less weight to the orbits of the 
outer, nearly equidistant 
planets which, in effect, determine the truncation radius. 
As a result, the optimization procedure also finds additional "near--minima" 
of high quality ($\sigma = 4.1$\%--4.7\%), among which the most extreme model has
the following parameter values: 
~$\delta = 2.5040$, ~$\beta_0 = 0.3806$, ~$R_0 = 0.02032$~AU, 
~$x_1 = 40$ ($:= 0.81$~AU), and ~$x_2 = 575$ ($:= 11.7$~AU).
A comparison between these values and the best-fit values listed in eq.~(\ref{neb}) gives 
us an idea about how shallow the region around the true minimum is in the four--parameter space
of the model. Note, in particular, that the power--law index does not differ 
from ~$\delta = 2.5$ ~by more than 1.5\% and the rotation parameter does not differ
from ~$\beta_0 = 0.4$~ by more than 5\% in any of the high--quality fits to the data.

The cylindrical Lane--Emden equation that we have solved can serve as a model of 
differentially--rotating disks supported by thermal pressure in the $z$--direction,
so we expect that the scale height from pressure support will be ~$H\propto R$ ~down 
the radial density gradient.\footnote{The vertical scale height of a pressure--supported 
disk is ~$H \sim c_0 /\Omega \sim R c_0 /v$,~
where $v$ is the rotation speed. In our models, $v$ is asymptotically flat when the density
exhibits a power--law profile and $c_0$ is constant, leading thus to the approximate relation 
~$H\propto R$. ~On the other hand, the scale height is 
~$H \sim c_0 /\Omega_0 = \sqrt{2} R_0 / \beta_0 = 3.4 R_0$ ~in the core of our best--fit
model. As a result, the mass estimates calculated in \S~\ref{md} for a disk with ~$H = R_0$ ~are 
too conservative.} In this case, the 
value of ~$\delta = 2.5$ ~implies that the corresponding power--law index in the surface--density 
profile ~($\Sigma\propto R^{-\delta + 1}$) ~of the nebular disk was ~$k = 1 - \delta = -1.5$. 
This value is virtually identical to that obtained by Weidenschilling (1977) who 
derived an estimate of  the surface density distribution of the protoplanetary disk 
by smearing out the observed planetary masses over annular rings in the disk's midplane 
and then applied a correction to this distribution by adding the appropriate amount of 
volatiles in order to bring the elemental abundance of the gas up to solar composition.
Our analytical work and modeling effort also provide a direct approach to the same problem,
but from a different angle than that conceived by Weidenschilling. The unambiguous congruence 
of the results obtained by these two disparate methods is rather astonishing
and suggests strongly that the surface density profile ~$\Sigma (R)$ ~of the solar nebula 
was indeed exhibiting an ~$R^{-1.5}$ ~power law in the isothermal phase of its evolution. 

Furthermore, our work also helps in delineating the fundamental physics behind such a mean surface 
density profile in the midplane of the solar nebula. With the aid of our best-fit model,
we can deduce substantial new information concerning the structure and 
the dynamics of the nebular disk. In addition to the structural and rotation parameters 
discussed above, we can use our analytic baseline model in order to probe 
the dynamical state of the protoplanetary disk in its isothermal phase as follows.

\subsubsection{Equation of State}\label{eos}

Using the length scale of the disk ($R_0=0.022440$~AU) 
in eq.~(\ref{length}), we can write an equation of state of the form
\begin{equation}
\frac{c_0^2}{\rho_0} \ = \ 4\pi G R_0^2 \ = \ 9.45\times 10^{16} \
{\rm ~cm}^5 {\rm ~g}^{-1} {\rm ~s}^{-2}\, ,
\label{crho}
\end{equation}
where $c_0$ and $\rho_0$ may be thought of as the local sound speed and the local density
in the inner disk, respectively.
For an isothermal gas at temperature $T$, ~$c_0^2 = {\cal R} T/\overline{\mu}$, 
~where $\overline{\mu}$ is the mean molecular weight and ${\cal R}$ is the
universal gas constant. Hence, eq.~(\ref{crho}) can be written
in the form
\begin{equation}
\rho_0 \ = \ 8.80\times 10^{-10}\left(\frac{T}{\overline{\mu}}\right) \
{\rm ~g} {\rm ~cm}^{-3}\, ,
\label{trho}
\end{equation}
where $T$ and $\overline{\mu}$ are measured in degrees Kelvin and 
${\rm ~g} {\rm ~mol}^{-1}$, respectively. 

For a cold disk of gas with $T = 10$~K 
and $\overline{\mu} = 2.34 {\rm ~g} {\rm ~mol}^{-1}$ (molecular hydrogen and
neutral helium with fractional abundances $X=0.70$ and $Y=0.28$ by
mass), we find that
\begin{equation}
\rho_0 \ = \ 3.76\times 10^{-9} \ {\rm ~g} {\rm ~cm}^{-3}\, .
\label{therho}
\end{equation}
This value is comfortably larger than the well-known threshold for planet formation in
the solar nebula ($\rho_*\approx 10^{-9} {\rm ~g} {\rm ~cm}^{-3}$; see
e.g., Lissauer [1993]) and implies that the conditions for planet formation
were already in place, at least in the inner disk, in the isothermal phase
of the disk's evolution.

\subsubsection{Rotational State}\label{rs}

Using the characteristic density of the inner disk
(eq.~[\ref{therho}]) in the definition of ~$\Omega_J\equiv\sqrt{2\pi G\rho_0}$, ~we
can determine the Jeans frequency of the disk:
\begin{equation}
\Omega_J \ = \ 3.97\times 10^{-8} {\rm ~rad} {\rm ~s}^{-1}\, .
\label{thej}
\end{equation}

Then, using the value $\beta_0 = 0.41465$ (eq.[\ref{neb}]) in the definition 
of ~$\beta_0\equiv \Omega_0 /\Omega_J$, we can determine the angular velocity
of the uniformly--rotating core ($R_1\leq 0.81931$~AU):
\begin{equation}
\Omega_0 \ = \ 1.65\times 10^{-8} {\rm ~rad} {\rm ~s}^{-1}\, .
\label{theom}
\end{equation}
For reference, this value of $\Omega_0$ corresponds to an orbital period 
of ~12~yr. In the {\it present solar system}, that would correspond to
a Keplerian orbit with semimajor axis ~$a = 5.24$~AU. So the core of the nebula
was rotating about as slowly as Jupiter is currently revolving around the Sun.

Finally, the constant asymptotic value of the angular velocity ~$\Omega_\infty$ ~for
~$x >> x_2$ ~is also a characteristic rotational parameter of the nebula (because the 
outer region of uniform density is necessary for the formation of the nearly equidistant
outer planets). Using eqs.~(\ref{rot}) and~(\ref{fasym}), we find that
\begin{equation}
\Omega_\infty \ \equiv \ \Omega_0 \left(\frac{x_1}{x_2}\right)^{\textstyle\delta /2} 
\ = \ 5.89\times 10^{-10} {\rm ~rad} {\rm ~s}^{-1} \, ,
\label{ominf}
\end{equation}
where the values listed in eq.~(\ref{neb}) ~and eq.~(\ref{theom}) were used in the 
numerical evaluation. This value of $\Omega_\infty$ is 28 times smaller than $\Omega_0$;
and corresponds to an orbital period 
of ~338 yr ~and to a Keplerian orbit with ~$a = 48.5$~AU ~near the outer edge of
the classical Kuiper belt in the present solar system.

\subsubsection{Mass Distribution}\label{md}

Assuming that our model disk is uniform in the $z$--direction
over a thin layer of thickness ~$2R_0$, ~we can multiply its mass per unit length
by this factor and we can define the total mass near the midplane of the solar
nebula out to a maximum radius $R_{max}$ as
\begin{equation}
M \ \equiv \ 2 R_0 \ \cdot \int_0^{R_{max}}{2\pi\rho(R) R dR} \ = \ 
2 M_0 \ \cdot\int_0^{x_{max}}{\tau (x) x dx} \, ,
\label{mass1}
\end{equation}
where the constant $M_0$ is given by
\begin{equation}
M_0 \ \equiv \ 2 \pi \rho_0 R_0^3 \ = \ 4.47\times 10^{-7} \ M_\odot \, ,
\label{mass2}
\end{equation}
where eq.~(\ref{therho}) ~and ~$R_0 = 0.022440$~AU ~were used in the numerical
evaluation. ~$M_0$ is approximately
the mass contained to within one length scale from the center of the disk
and implies that the central surface density is
\begin{equation}
\Sigma_0 \ \equiv \ \frac{M_0}{\pi R_0^2} \ = \ 2 R_0\cdot\rho_0 \ 
\approx \ 2520 \ {\rm ~g} {\rm ~cm}^{-2} \, .
\label{sig}
\end{equation}
Using this value, the core radius $R_1=0.81931$~AU, and the power--law index 
$k\approx -1.5$, we can write the surface density profile of the solar nebula
for $R_1\leq R\leq R_2$ in the form
\begin{equation}
\Sigma (R) \ = \ 1870 \left( \frac{R}{1 ~{\rm AU}}\right)^{-1.5} \ \ 
{\rm ~g} {\rm ~cm}^{-2} \, ,
\label{sig2}
\end{equation}
where $R$ is measured in~AU.
The value at 1~AU is lower than that estimated by Weidenschilling (1977);
but it agrees very well with Hayashi's (1981) competing result that was obtained 
by the same method and with the same data, and also with Kuchner's (2004) 
average estimate obtained from a similar analysis of 26 multiple--planet 
extrasolar systems.

Adopting now the baseline density profile (eq.~[\ref{den3}]) as an approximation to the actual
density distribution and using the parameters listed in eq.~(\ref{neb}), we can evaluate
the integral of eq.~(\ref{mass1}) over the three regions of the baseline out to, e.g.,
$R_{max}=50$~AU ~($x_{max}=2228$):
\begin{eqnarray}
M \ & = &  2 \ M_0 \ \beta_0^2 \ \cdot
\left[ \ \int_0^{x_1}{x dx} 
   \ + \ \int_{x_1}^{x_2}{\left(\frac{x_1}{x}\right)^{\textstyle\delta} x dx} 
   \ + \ \int_{x_2}^{x_{max}}{\left(\frac{x_1}{x_2}\right)^{\textstyle\delta} x dx} 
   \right] \nonumber \\
 & & \nonumber \\
    & = & \ \ \ \ \ \ \ \ \ \ \ 
                            \ \ \ 1\times 10^{-4} \ M_\odot 
                        \ \ + \ \ 3\times 10^{-4} \ M_\odot 
                        \ \ + \ \ 5\times 10^{-4} \ M_\odot \\
% & & \nonumber \\
    & \approx & \ \ \ \ \ \ \ \ \ \ \ 
                            \ \ \ 0.001 \ M_\odot \nonumber \ .
\end{eqnarray}
The total mass is one order of magnitude smaller than the low end of the estimate 
for the "minimum--mass solar nebula" ($0.01~M_\odot$; ~Weidenschilling 1977).
This is not surprising since we have adopted a very small value ($R_0$) for the
vertical scale of the disk. In the classical scenario of cloud collapse, this mass near
the midplane will be enhanced substantially as more matter from the
parent cloud will continue falling onto the disk (see \S~\ref{ipc} below for more details).
It is however interesting that the inner disk already has high enough densities 
(eq.~[\ref{therho}]) to begin the process of planet formation so early in its evolution, 
before the central protostar grows to become gravitationally dominant or radial accretion
becomes important in the gas.

\subsubsection{Integral Properties of the Core}\label{ipc}

Integrating over the mass distribution 
of the core of the baseline model, we can estimate important dynamical properties,
such as the core mass and angular momentum. The analytic estimates can then be compared
to the corresponding results from protostellar collapse simulations. Such 
simulations have been recently reviewed by Tohline (2002) who also summarized
the main results over which there seems to be widespread agreement among researchers 
working in the field for almost 40 years (see the discussion of Tohline centered around his 
Figure 2 and Table 2). 

The mass of the core found in \S~\ref{md} above, ~$M_1\approx 10^{-4} \ M_\odot$, ~is 40 times smaller 
than the typical mass of a collapsing cloud core in which the Jeans instability may be 
temporarily halted by thermal pressure and an adiabatic 
quasi--equilibrium may then be established (see case B in Table 2 of Tohline [2002]). This 
means that the core of our quasi--equilibrium disk will remain isothermal and it will 
continue to accumulate mass from the infalling cloud beyond the point described here.

Furthermore, the total angular momentum of the core of the 
model is
\begin{equation}
L_1 \ \equiv \ 2 \ R_0 \ \cdot\int_0^{R_1}{\Omega_0 R^2 \cdot 2\pi\rho_0 R dR} \
= \ 2 M_0 \Omega_0 R_0^2 \ \cdot \int_0^{x_1}{x^3 dx} \ 
= \ \frac{1}{2} M_0 \Omega_0 R_0^2 x_1^4 \, .
\label{angmom1}
\end{equation}
This implies that the {\it specific} angular momentum of the matter in the core is
\begin{equation}
\frac{L_1}{M_1} \ = \ \frac{\Omega_0 R_0^2 x_1^2}{2 \beta_0^2} \
= \ 7.19\times 10^{18} {\rm ~cm}^2 {\rm ~s}^{-1} \, ,
\label{angmom2}
\end{equation}
which is about one--half of the corresponding estimate given 
by Tohline (2002) for a cloud core at the endpoint of its isothermal evolution. 
Clearly, in a realistic setting, this type of low--mass core has the potential to grow 
by accreting a lot more matter of low specific angular momentum, the 
kind that can settle on to the central region from the vertical direction.

\section{Summary and Discussion}\label{discussion}

\subsection{Summary}

We have presented exact analytic and numerical solutions of the axisymmetric,
cylindrically symmetric, isothermal Lane--Emden equation with and without rotation
(eqs.~[\ref{main3}] and~[\ref{main2}], respectively). This second--order ODE 
describes the radial equilibrium configurations that are available to a 
self--gravitating perfect fluid in which the {\it vertical variation}
of the enthalpy gradient, namely
$$
\frac{d}{dz}\left(\rho^{-1}\frac{dP}{dz}\right) \ ,
$$
is negligible. The enthalpy gradient $\rho^{-1} dP/dz$,
where $\rho$ is the density and $P$ is the pressure, is the term
that establishes vertical hydrostatic equilibrium when it
successfully balances the gravitational force $d\Phi /dz$ in the
$z$--direction, but it is its vertical variation that is actually
ignored when the cylindrical Lane--Emden equation is considered. This
term is zero identically for a cylindrical model of
infinite vertical extent; such a model may be applicable to
elongated, filamentary, star--forming regions in which $z > R$. In
other astrophysical applications, especially those dealing with
gaseous disks that concern us in this work, this second derivative
is not identically zero everywhere; but it is ignored on the basis
that it vanishes on the symmetry plane and hopefully it remains
negligible away from that plane over some layer
of the astrophysical disk (as in the exact models of Schmitz \& Ebert [1986]).
In the specific case of the
solar nebula, this assumption is probably valid because the
disk is not in vertical hydrostatic equilibrium 
and vertical pressure support is smaller than
the weight of the infalling matter everywhere,
except near the midplane of the disk where the 
gradients tend to zero by symmetry anyway.

The rotating analytic solutions are exact intrinsic solutions of the Lane--Emden
equation~(\ref{main3}). Their density profiles are all pure power laws or combinations
of power laws (eqs.~[\ref{class2}] and~[\ref{den3}]). These solutions are determined
solely by the balance between the {\it gradients} of the centrifugal and the gravitational
accelerations (eq.~[\ref{gradients}]) because for an isothermal fluid with a power--law
density profile, the {\it radial variation} of the enthalpy gradient, namely
$$
\frac{d}{dR}\left(\rho^{-1}\frac{dP}{d\ln R}\right) \ ,
$$
is zero identically (see eq.~[\ref{part2}] and the discussion at the end of \S~\ref{withrot}).
The differential--rotation profiles
that can support such power--law densities are quite general and varied
(eqs.~[\ref{class3}] and~[\ref{class4}]). They contain three integration constants that can be used
to compose models of equilibrium disks in which the density profiles are nonsingular and arbitrarily
steep in radius. We have created one such composite equilibrium model and we have
applied it to the structure of the midplane of the solar nebula (\S~\ref{solar}).
In this model, we have connected a central core region of constant density and an outer region of
constant density with a power--law profile. The rotation parameter and the size of the core as
well as the power--law index and the truncation radius of the intermediate region 
are the four model
parameters to be determined based on the presently observed solar system. Beyond the truncation
radius, the density remains constant at a low level. This turns out to be the necessary nebular
background for the formation of nearly equidistant outer planets and its physical origin deserves
further investigation.

The density profile of the composite equilibrium disk model is no longer an exact solution when physical
boundary conditions are imposed at the center of the disk. The exact solution to the
associated
boundary--value problem is obtained by numerical integration that enforces the proper central
boundary conditions in a model with the same differential--rotation law as the composite analytic model.
We have found that the intrinsic analytic density profile is a good approximation to the actual
numerical solution, i.e., it is a ``baseline" that exhibits on average all the important
features of the exact solution to the boundary--value problem, except one: the actual density profile is
permanently oscillatory in radius. In fact, it oscillates around the baseline solution because it is
attracted to this inherent particular solution of the Lane--Emden equation, but the two solutions
cannot match in any finite segment since such coincidence is strictly prohibited by the imposed
central boundary conditions (see Fig.~2 and the discussion in \S~\ref{compo}).

The radial density peaks of the actual solution to the
boundary--value problem correspond to local minima of the
gravitational potential in the midplane of the protoplanetary disk.
In a cloud collapse scenario, the density of the gas and the
concentration of condensates (dust, rocks, and ices) will be
enhanced inside these local gravitational potential wells as more
matter rains down onto the disk from the surrounding protostellar
cloud. It is not unreasonable to consider that the added
material will, in turn, deepen further these potential wells which
may thus become feasible sites of planet formation. Based on this
picture, we have proceeded to fit the density peaks of the exact
numerical solutions of the Lane--Emden boundary--value problem to
the presently observed locations of the planets around the Sun and
the satellites of the Jovian planets. Our best--fit model for the
planets in our solar system (Fig.~3 and Tables~2 and ~3) is
described in \S~\ref{solar2} and the dynamical parameters obtained
for the solar nebula are described in \S~\ref{solar3}. The mean
relative error for this best--fit model is 4.1\% and the relative
deviations for individual planets do not exceed 6.5\%. Furthermore,
this model suggests that the radial surface density profile of the
protoplanetary disk between roughly 0.8~AU and 11.3~AU declined as
$R^{-1.5}$, a result that agrees fully with the classic
determinations of Weidenschilling (1977) and Hayashi (1981) 
which were based on the
presently observed planetary masses. We find the confluence
of these very diverse approaches encouraging and supportive of the
view that the fundamental notions behind our model are indeed
relevant to the problem of the radial distribution of planetary orbits. 
%We believe that further consideration along the same lines (e.g. simulations) 
%could help us understand better the conditions that prevailed during the
%early evolution of the solar nebula. 
We would also like to note
that the same type of modeling works very well for the regular satellites
of Jupiter and for the 5 planets of 55 Cancri, 
and we plan to present these results in future publications.

\subsection{Discussion}\label{dis}

{ Our work is the first to produce exact solutions of the nonlinear isothermal 
Lane--Emden equation with differential rotation that exhibit a pronounced oscillatory 
behavior, a 
feature attributed exclusively to the proper choice of the central
boundary conditions. The roughly arithmetic or geometric spacings of the
corresponding density maxima depend on the local slope of the differential rotation
profile and provide a transparent, physically--based,
and soundly formulated reason for producing a sequence of distinct
density enhancements at radial positions that agree with the present
planetary distances, given that the latter are observed to
follow an arithmetic, then a geometric, and finally again an arithmetic
progression (see Table 1). Hence, it should be remarked that we now have at
hand, for the first time, a plain explanation of the so-called 
Titius--Bode ``law" of planetary distances.
%
%The physical understanding gained from our analytic calculations and modeling effort
%provides a straightforward resolution of the centuries--old controversy surrounding the
%so--called Titius--Bode ``law" of planetary distances
%(see Table 1).
This empirical algorithm has been known for 240 years but with
no underlying physical justification} (see e.g. Graner \& Dubrulle
[1994] and Hayes \& Tremaine [1998]). Our best--fit model of the
midplane of the solar nebula (Fig.~3 and Table~2) shows {
a progression of density enhancements
% attempts to match the locations of the planets that formed along the steep gradient of the
%mean (baseline) density profile. As can be seen in Fig.~3a, these planets have
%formed at density peaks
that are stretched farther apart from one another, as the actual
density profile decreases sharply in trying to keep up with the
steeply declining baseline. The relative spacing of the peaks is
roughly geometric between 0.7 and 20~AU, and this explains the success of 
the Titius--Bode
algorithm for a large number of planetary orbits. { The
change of the underlying density profile to constant beyond a
certain distance in our model can then also explain why this
algorithm fails at large distances. A similar account of the
nongeometric spacing of Mercury also finds a straightforward
explanation in terms of a uniform central core in our model.}
%The rigid geometric spacing of the algorithm cannot however account for
%the approximately flat sections of the density profile in the inner
%core and in the outer disk, and this is precisely where the ``law"
%has been known to fail. The problem in the inner solar system was
%remedied many years ago by the insertion of a nongeometric term in
%the beginning of the ad--hoc geometric progression, but there is no
%justifiable remedy for tapering off the Titius--Bode geometric
%progression at the other end (in order to capture the arithmetic
%spacing observed in the locations of the outer planets), and the
%algorithm is left to fail beyond the orbit of Uranus.
So, just as many researchers have suspected in the past (see \S~\ref{TB}), 
the Titius--Bode algorithm has no underlying physical
principle underneath its phenomenology; the
algorithm has been successful only to the extent that it has managed
to exploit empirically the most pronounced feature of the radial
density profile of the solar nebula (the roughly geometric
spacing of the peaks between 0.7 and 20~AU). Finally, we should point out
that the above resolution of the long--standing problem
of planetary "order" in our solar system is very anticlimactic, as}
the long--sought physical explanation turns out to lie entirely within
the realm of conventional physics and does not need to invoke unrestrained 
numerology of the Titius--Bode  type, exotic physics, "new" dynamical laws, or
arbitrary "universal" constants and ad--hoc solar--system
"quantizations" such as those proposed in the past.

{ We view this work only as a first (but compelling) step 
toward understanding the systematic formation of planets; the evolution 
that results in the formation of the entire solar system 
likely involves additional processes and physics (e.g.,
dissipative disk accretion, magnetic fields) that are necessary 
in delineating its present structure and composition. 
However, we believe that the condensed solid cores inside the
gravitational potential minima of the solar nebula could possibly
survive the subsequent evolution and could form planets at the same
orbital locations. 
Furthermore, we hope that the results presented here
will provide the motivation behind the formulation of some }
%
%The results presented in this work may be used to formulate some
novel hypotheses concerning the formation
and the dynamical evolution of certain sectors in the early solar system
and in the
recently discovered extrasolar planetary systems (see the reviews of Lissauer
[1993], Beckwith \& Sargent [1996], and Marcy et al. [2005]).
Some of the ideas that emerge from our study are outlined below.

\subsubsection{The Distribution of Condensates in the Protoplanetary Disk}

Our results support the
segregation of condensates at specific radii in which the gas density is larger than the mean
background value (Fig.~3 and Table~2). These orbital locations in the midplane of the disk are ideal sites
for growing planitesimals by accumulation of smaller bodies in a systematic (nonrandom) way.
In the best--fit density profile $\tau\propto x^{-\textstyle\delta}$ with $\delta = 2.5$,
the absolute spacing of these sites of concentrated material 
depends on a single parameter, the length scale of the disk
($R_0 = 0.022440$~AU) which is a measure of the entropy content of the gas (see eq.~[\ref{crho}]).

The segregation of condensates in initially shallow (see Table 3) gravitational potential wells should be
tested by computational experiments because this kind of work may potentially lead to improved protoplanetary
models. Numerical work is quite common in investigations of the early
solar system but the simulations have always started from arbitrary initial conditions
(e.g., Wetherill 1989; Ruzmaikina, Safronov, \& Weidenschilling 1989; Bodenheimer, Ruzmaikina, \&
Mathieu 1993; Boss 1995) because of the limited information that can be collected from observations,
meteorites, and laboratory experiments. Our results can potentially provide 
a better handle on the appropriate initial conditions for simulations of this type.

\subsubsection{Planet Migration in Our Solar System}

Numerical investigations of planet migration have
intensified over the past 10 years (see the progress report by Levison et al. [2007]) in an attempt to
understand the complex dynamics observed in the Kuiper belt: the trapping of some small objects in the
mean--motion resonances of Neptune, the gravitational scattering of other small objects to high
inclinations, the severe mass deficiency (by at least a factor of 100) of the entire Kuiper belt,
and the abrupt outer ``edge" or ``gap" of the classical Kuiper belt at 47--48~AU (e.g., Delsanti \&
Jewitt 2006). Our results do not support the idea that the gaseous giant planets have secularly
migrated over time to the presently observed radial locations (Fern\'andez \& Ip 1984;
Malhotra 1993, 1995; Gomes, Morbidelli, \& Levison 2004;
Gomes et al. 2005). The quality of our best--fit model of planetary distances is so high
(see \S~\ref{solar2}) that it seems improbable that any planet that scattered small planitesimals has
managed to jump out of its local gravitational potential well and has moved to a new location inside
an adjacent potential well.

On the other hand, the displacement of a planet within its potential well appears viable
in the context of our model, and the potential wells of the gaseous giants are quite large in
size (see Table~3 and Fig.~3a): 
Uranus and Neptune are both orbiting on potential minima of radial half--width 
$\Delta d\approx\pm 5$~AU; 
Saturn's minimum has $\Delta d\approx +5/-0.06$~AU;
and Jupiter's minimum has $\Delta d\approx +5/-0.4$~AU. 
These values are not very different than
the migration distances used so far in the numerical simulations; 
only Saturn presents a challenge because its potential well 
is too steep at smaller radii to allow for an outward 
migration of more than $\sim$0.1~AU to its present location. 
Perhaps new models of planet migration can be constructed in which 
these results will be taken into account.

\subsubsection{An Inner Gap in Protostellar Disks}

Fig.~3b and Table 3 show that the first density minimum
in the best--fit model of the solar nebula occurs at $d = 0.179$~AU. This suggests that there were
no likely planetary sites interior to the orbit of Mercury. It also suggests that there was a
significantly lower concentration of condensates within this minimum, in the area below
the baseline that extends
approximately from 0.1~AU to 0.3~AU. Since we expect that our solar system is in no way special
but rather representative of large planetary systems around solar--type stars,
we believe that the same deficiency of solids should also exist in the inner regions
of other protostellar disks that are currently in the process of forming their protostars and protoplanets.
We think that such an inner gap in condensates is currently inferred in the circumstellar disks of some
pre--main--sequence stars; these disks show almost no near--infrared radiation emanating from the inner
$0.2-0.3$ AU (Strom, Edwards, \& Skrutskie 1993; Beckwith \& Sargent 1993; 
Millan--Gabet et al. 2001; Akeson et al. 2005; Fedele et al. 2008).
The alternative view is that an orbiting protoplanet may have cleared a gap in that area.
Although this may be possible in small heavy disks where large planets could form very close to 
their stars (Christodoulou \& Kazanas 2008), it cannot explain large systems like our own 
in which gas giants are not expected to form within the inner 0.3~AU.

\subsubsection{Extrasolar Planetary Systems}

It is believed on theoretical grounds that terrestrial planets
are commonly formed near their central stars and that it is unlikely to have a gaseous giant form closer
to the center than about 4~AU (Boss 1995). Accordingly, our solar system is presumed to be typical
of planetary systems around other stars. On the other hand, most of the extrasolar planets
found to date do not fit in this theoretical picture. The currently available
sample of extrasolar planets is strongly biased toward giant planets because small perturbing masses
cannot be easily resolved by observations,
so it is understood 
that it will take more time before large extrasolar systems comparable to our own
can be detected. 
Nevertheless, there is presently no doubt that in
many extrasolar systems, a large number of massive planets exist very close to their stars
(the trail of discovery started with the planet at 0.05~AU in 51~Peg found by Mayor \& Queloz [1995]
and the planet at 0.48~AU in 70 Vir
found by Marcy \& Butler [1996]). This discrepancy seems to point to the hypothesis that the cores of these
planets have formed farther away and that the objects have migrated inward to the observationally inferred
distances (Kary \& Lissauer 1995; Lin, Bodenheimer, \& Richardson 1996), perhaps destroying in the process
the orbits of terrestrial planets that could have existed closer to central stars.

Migrating giant planets cannot be accounted for in our model,
so we will have to wait until observations can detect some
large extrasolar systems similar to our own.
However, two small multiple--planet systems have already been detected
which are different than our solar system, but their planetary
distributions appear to be more in line 
with a sequence of well--ordered density peaks as predicted by our model.
These systems are 55~Cancri (Fischer et al. 2008) and
HD~37124 (Vogt et al. 2005).
Using updated data from the "Catalog of Nearby Exoplanets"
(http://exoplanets.org and Butler et al. [2006]),
we find that the ratios of semimajor axes of the three innermost planets 
in these two systems are
~$a_2/a_1 \simeq 3$ ~and ~$a_3/a_2 \simeq 2$. These ratios are barely larger
than those in our solar system (1.9 and 1.4, respectively).
~In contrast, all the other systems in the Catalog with three or more exoplanets 
have ~$a_2/a_1 > 6$~ and/or ~$a_3/a_2 > 3$, ~ratios that are too large compared
to those in 55~Cancri, HD~37124, and our solar system.

We see then that the relative distributions of planetary orbits
in 55~Cancri and HD~37124 are effectively the same, despite the fact that
these two well--ordered systems are much smaller than our solar system
and their central stars have widely different chemical compositions
(55~Cancri is very metal--rich, while HD~37124 is very metal--poor).
55~Cancri is now believed to have 5 planets in remarkably circular orbits 
(Fischer et al. 2008), and this number is enough to allow for our type of modeling
for this system. We have carried out the analysis of 55~Cancri and we
present our results in a companion paper (Christodoulou \& Kazanas 2008).

%\acknowledgments


\newpage

\section*{FIGURE CAPTIONS}

\figcaption{Analytic density and rotation profiles of the composite 
equilibrium models described by eqs.~(\ref{den3}) and~(\ref{den31})
for $x_1=100$, $x_2=500$, and $\delta =$ 2.5, 3, and 4. The density
profile $\tau_{base} (x)/\beta_0^2$ is uniform for $x\leq x_1$ and for $x\geq x_2$; and
it follows the power law ~$x^{-\textstyle\delta}$ ~in the in--between region.
The rotation profile $f(x)$ is uniform for $x\leq x_1$ and monotonically
decreasing for $x > x_1$, as specified by eq.~(\ref{den31}); at very large
radii (for $x >> x_2$), $f(x)$ approaches the constant asymptotic value 
~$(x_1/x_2)^{\textstyle\delta /2}$ .
\label{fig1}
}

\figcaption{Equilibrium density profile for a model with rotation parameter $\beta_0 = 0.2$ 
and a composite rotation profile given by eq.~(\ref{den31}) with ~$x_1=200$, ~$x_2=1000$, ~and
~$\delta = 3$. Frame (a) shows the radial distance ~$x$ ~on a linear scale out to ~$x=2000$. 
Frame (b) shows the same radial distance on a logarithmic scale out to ~$\ln x=8$ ~($x=2981$). 
The physical density 
$\tau (x)$ (solid line) satisfies the boundary conditions~(\ref{bc}) and, as a result, 
it is forced to oscillate permanently about the inherent baseline solution $\tau_{base} (x)$ 
(eq.~[\ref{den3}], dashed line) of the Lane--Emden ODE~(\ref{main5}). The nonrotating 
analytic solution (eq.~[\ref{SO}], dash-dotted line) is also shown for reference.
\label{fig2}
}

\figcaption{Equilibrium density profile for the midplane of the solar 
nebula. The composite model described in \S~\ref{solar1} (eq.~[\ref{den3}], dashed line) 
has been adopted for the RHS of the Lane--Emden equation~(\ref{main5}) and this ODE has 
been integrated numerically subject to the physical boundary conditions~(\ref{bc}). The 
resulting solution (solid line) has been fitted to the present solar system so that its 
density maxima (dots) correspond to the observed semimajor axes of the planetary orbits 
(open circles).
The third density maximum is always scaled to a distance of 1~AU in this procedure; in the
best--fit model shown here, it occurs at $x=44.564$ and implies that the length scale of the 
solar nebula was $R_0 = 0.022440$~AU. The mean relative error of the fit is
~4.1\%, ~affirming that this simple equilibrium model produces an incomparable match to the 
observed data.
Frame (a) shows the radial distance ~$d$ ~on a linear scale out to ~$d=40$~AU. ~Frame (b) 
shows the same radial distance on a logarithmic scale out to ~$\ln d = 4.5$ ~($d=90$~AU).
~The nonrotating analytic solution (eq.~[\ref{SO}], dash-dotted line) is also 
shown for reference.
\label{fig3}
}

\newpage

\begin{figure}[t]
\vskip 7.5truein
\includegraphics{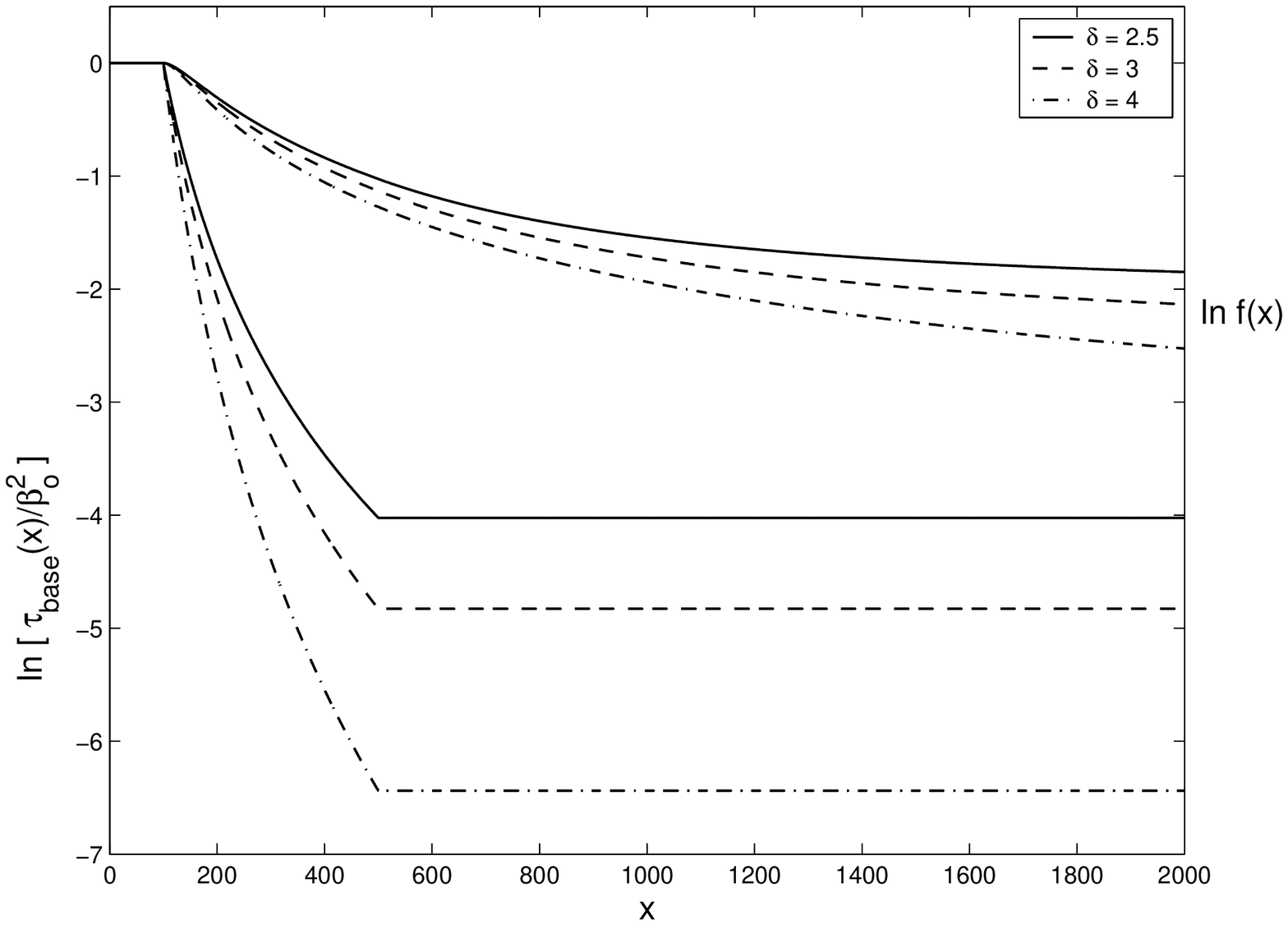}
{\bf FIGURE 1}
\end{figure}

\newpage

\begin{figure}[t]
\vskip 2.5truein
\includegraphics{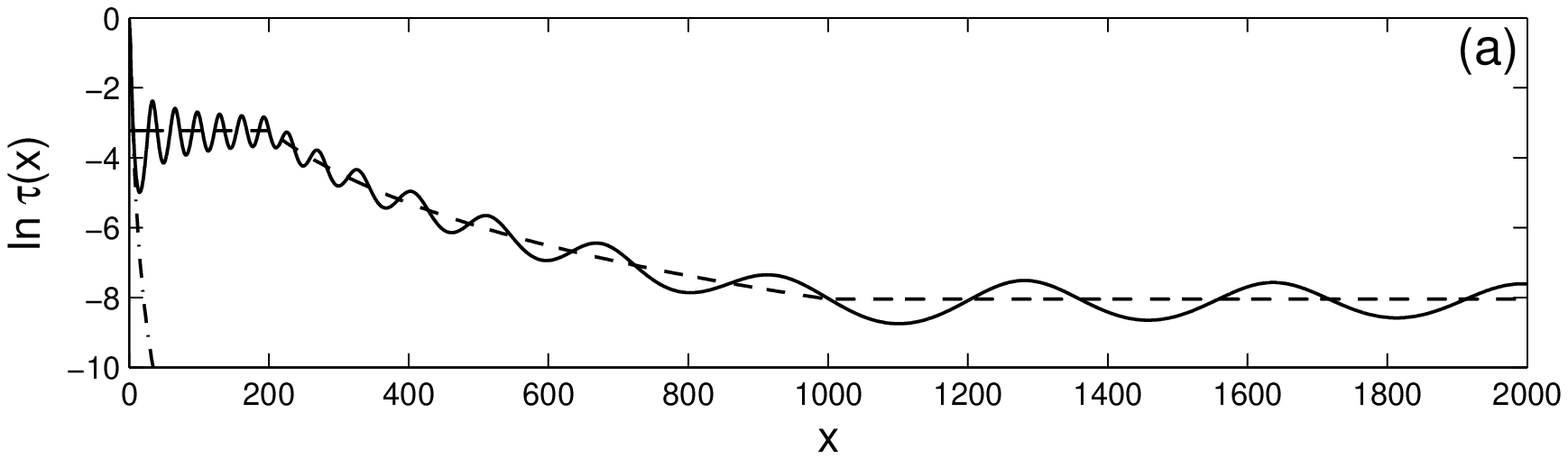}
\vskip 2.5truein
\includegraphics{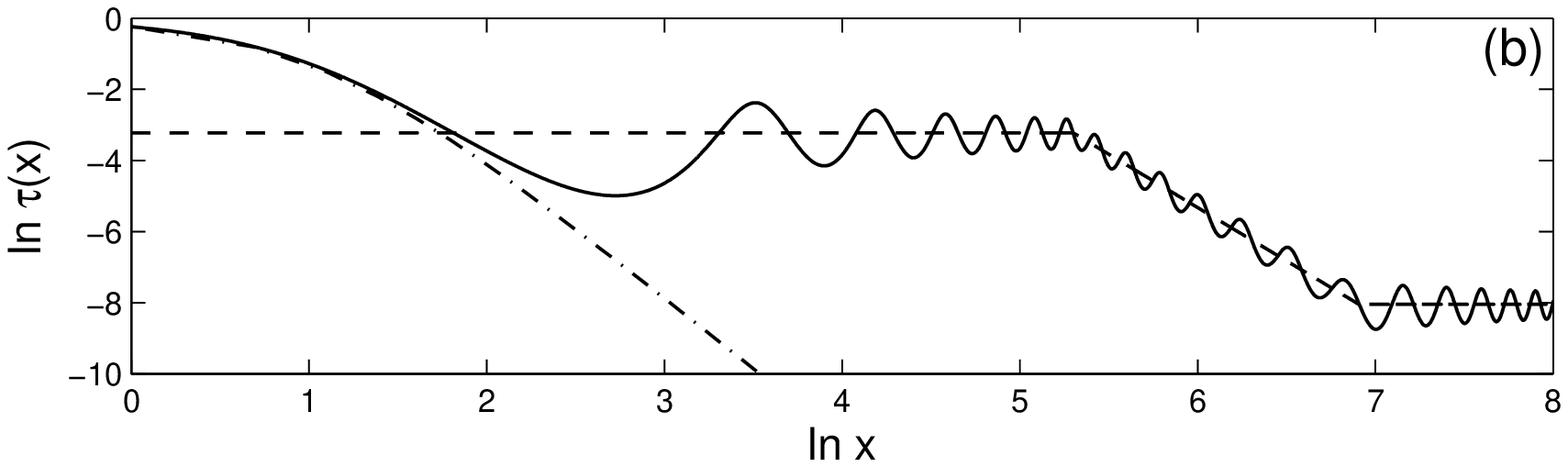}
{\bf FIGURE 2}
\end{figure}

\newpage

\begin{figure}[t]
\vskip 2.5truein
\includegraphics{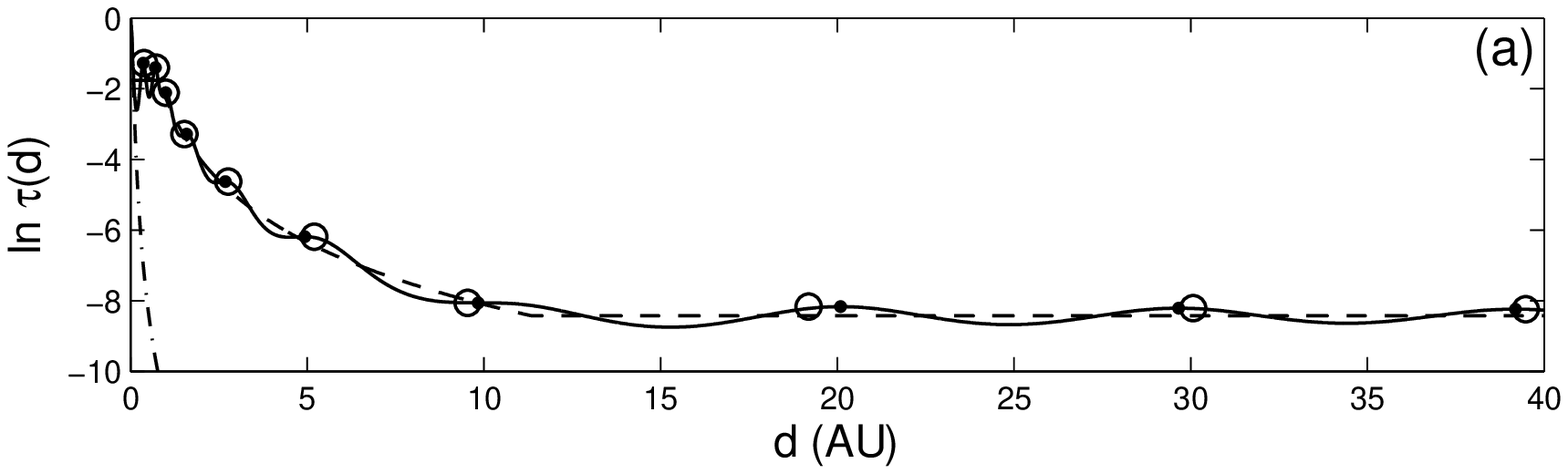}
\vskip 2.5truein
\includegraphics{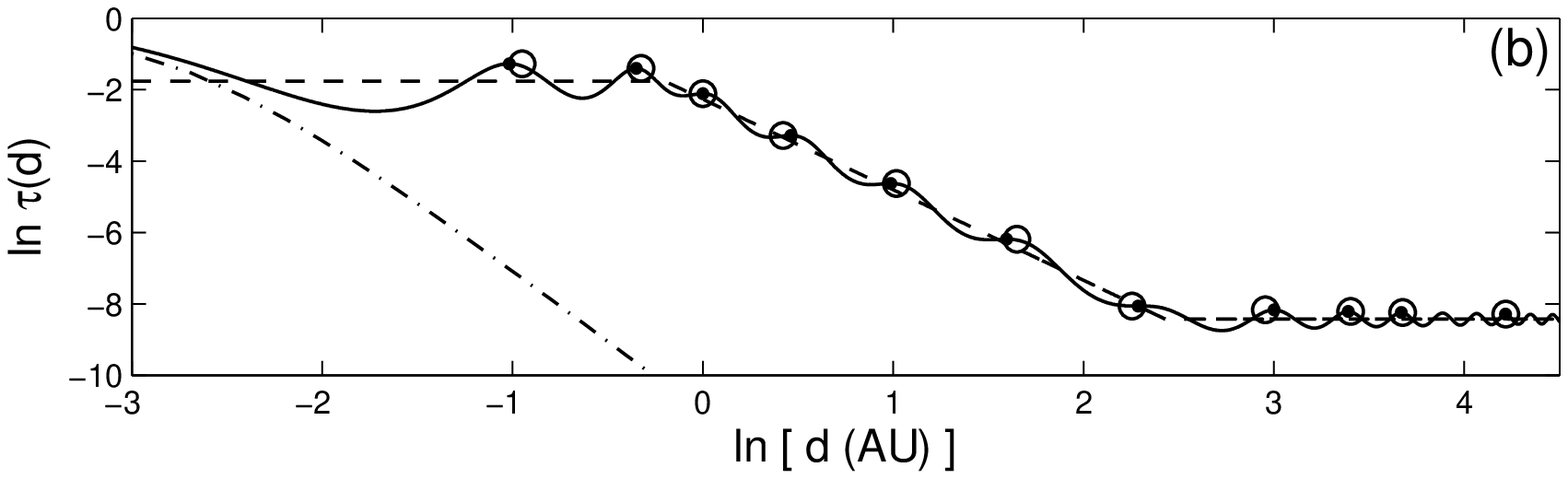}
{\bf FIGURE 3}
\end{figure}


\newpage
\begin{thebibliography}{}

\bibitem[1a]{1a}
Akeson, R. L., Boden, A. F., Monnier, J. D., et al. 2005, \apj, 635, 1173

\bibitem[Beckwith \& Sargent (1993)]{bs93}
Beckwith, S. V. W., \& Sargent, A. I. 1993, in Protostars
and Planets III, ed. E. H. Levy, \& J. I. Lunine (Tucson: Univ.
of Arizona Press), 521

\bibitem[2]{2}
Beckwith, S. V. W., \& Sargent, A. I. 1996, Nature, 383, 139

\bibitem[2aa]{2aa}
Bender, C. M., \& Orzag, S. A. 1978, Advanced Mathematical Methods for
Scientists and Engineers, (New York: McGraw--Hill)

\bibitem[2b]{2b}
Berkovich, L. M. 1997, Symm. Nonlinear Math Phys, 1, 155

\bibitem[2a]{2a}
Binney, J., \& Tremaine, S. 1987, Galactic Dynamics 
(Princeton: Princeton Univ. Press)

 %\bibitem[3]{3}
 %Binzel, R. P. 1989, in Asteroids II, ed. R. P. Binzel, T. Gehrels, \&
 %M. S. Matthews (Tucson: Univ. of Arizona Press), 3

\bibitem[4]{4}
Bodenheimer, P., Ruzmaikina, T., \& Mathieu, R. S. 1993, in Protostars
and Planets III, ed. E. H. Levy, \& J. I. Lunine (Tucson: Univ.
of Arizona Press), 367

\bibitem[5]{5}
Boss, A. P. 1995, Science, 267, 360

\bibitem[5a]{5a}
Brown, M. E., Trujillo, C. A., \& Rabinowitz, D. L. 2005, \apj, 635, L97

\bibitem[6]{6}
Butler, R. P., \& Marcy, G. W. 1996, \apj, 464, L153

\bibitem[6b]{6b}
Butler, R. P., Wright, J. T., \& Marcy, G. W., et al. 2006, \apj, 646, 505

\bibitem[6a]{6a}
Chandrasekhar, S. 1939, An Introduction to the Study of Stellar Structure,
(Chicago: Univ. of Chicago Press)

\bibitem[7]{7}
Christodoulou, D. M. 1993, \apj, 412, 696

\bibitem[8]{8}
Christodoulou, D. M., Contopoulos, J., \& Kazanas, D. 1996, \apj, 462, 865

\bibitem[8b]{8b}
Christodoulou, D. M., \& Kazanas, D. 2008, \apj, submitted

\bibitem[9]{9}
Christodoulou, D. M., \& Narayan, R. 1992, \apj, 388, 451

\bibitem[9b]{9b}
Christodoulou, D. M., Shlosman, I., \& Tohline, J. E. 1995, \apj, 443, 551

\bibitem[10]{10}
Coxeter, H. S. M. 1989, Introduction to Geometry, 2nd edition (New
York: Wiley)

\bibitem[11]{11}
Danby, J. M. A. 1988, Fundamentals of Celestial Mechanics, 2nd edition
(Richmond: Willmann--Bell)

\bibitem[11a0]{11a0}
Delsanti, A., \& Jewitt, D. 2006, in Solar System Update, 
ed. Ph. Blondel \& J. W. Mason (Berlin: Springer--Praxis), 267

\bibitem[11a]{11a}
Dixon, J. M., \& Tuszy\'nski, J. A. 1990, Phys. Rev. A, 41, 4166

\bibitem[11b]{11b}
Dubrulle, B., \& Graner, F. 1994, A\&A, 282, 269

  %\bibitem[12]{12}
  %Duncan, M. J., \& Quinn, T. 1993, \araa, 31, 265

\bibitem[12a]{12a}
Emden, R. 1907, Gaskugeln (Leipzig: B. G. Teubner)

\bibitem[12a1]{12a1}
Fabbiano, G. 1989, \araa, 27, 87

\bibitem[12a0]{12a0}
Fabian, A. C. 1991, A\&A Rev., 2, 191

\bibitem[12b5]{12b5}
Fedele, D., van den Ancker, M. E., Acke, B., et al. 2008, A\&A, preprint (astro-ph/0809.3947)

\bibitem[12c1]{12c1}
Fermi, E. 1927, Rend. Accad. Naz. Lincei, 6, 602

\bibitem[12d1]{12d1}
Fern\'andez, J. A., \& Ip, W.--H. 1984, Icarus, 58, 109

\bibitem[182c]{182c}
Fischer, D. A., Marcy, G. W., Butler, R. P., et al. 2008, \apj, 675, 790

\bibitem[12b1]{12b1}
Fowler, R. H. 1914, Quart. J. Math., 45, 289

\bibitem[12b2]{12b2}
Fowler, R. H. 1930, \mnras, 91, 63

\bibitem[12c]{12c}
Fowler, R. H. 1931, Quart. J. Math., 2, 259

\bibitem[12c3]{12c3}
Frank--Kamenetskii, D. A. 1955, Diffusion and Heat Exchange in Chemical Kinetics
(Princeton: Princeton Univ. Press)

\bibitem[12c34]{12c34}
Goenner, H. 2001, Gen. Rel, Grav., 33, 833

\bibitem[12c4]{12c4}
Goenner, H., \& Havas, P. 2000, J. Math. Phys., 41, 7029

\bibitem[12c5]{12c5}
Gomes, R. S., Morbidelli, A., \& Levison, H. F. 2004, Icarus, 170, 492

\bibitem[13a0]{13a0}
Gomes, R. S., Gallardo, T., Fern\'andez, J. A., \& Brunini, A. 2005, 
Celest. Mech. Dyn. Astron., 91, 109

\bibitem[13]{13}
Goodman, J., \& Narayan, R. 1988, \mnras, 231, 97

 %\bibitem[14]{14}
 %Gradie, J. C., Chapman, C. R., \& Tedesco, E. F. 1989, in Asteroids
 %II, ed. R. P. Binzel, T. Gehrels, \& M. S. Matthews (Tucson:
 %Univ. of Arizona Press), 316

\bibitem[14a]{14a}
Graner, F., \& Dubrulle, B. 1994, A\&A, 282, 262

 %\bibitem[14b]{14b}
 %Gray, D. F. 1997, Nature, 385, 795

\bibitem[15]{15}
Hansen, C. J., Aizenman, M. L., \& Ross, R. R. 1976, \apj, 207, 736

\bibitem[16]{16}
Hayashi, C. 1981, Prog. Theor. Phys. Suppl., 70, 35

\bibitem[16a]{16a}
Hayes, W., \& Tremaine, S. 1998, Icarus, 135, 549

\bibitem[16b1]{16b1}
Horedt, G. P. 1986, A\&A, 160, 148

\bibitem[16b2]{16b2}
Jaki, S. 1972, Am. J. Phys., 40, 93

\bibitem[16c]{16c}
Jeans, J. H. 1914, Phil. Trans. Royal Soc. London, Series A, 213, 457

\bibitem[17]{17}
Kary, D. M., \& Lissauer, J. J. 1995, Icarus, 117, 1

\bibitem[18]{18}
Kaufmann, W. J., III 1994, Universe, 4th edition (New York: Freeman)

\bibitem[18a1]{18a1}
Kochanek, C. S. 1995, \apj, 445, 559

\bibitem[18a22]{18a22}
Kuchner, M. J. 2004, \apj, 612, 1147

\bibitem[18a]{18a}
Lane, L. J. H. 1869-70, Amer. J. Sci. Arts, 4, 57

\bibitem[18a2]{18a2}
Lecar, M. 1973, Nature, 242, 318

\bibitem[18a3]{18a3}
Levison, H. F, Morbidelli, A., Gomes, R. S., \& Backman, D. 2007,
in Protostars and Planets~V, ed. B. Reipurth, D. Jewitt, \& K. Keil (Tucson:
Univ. of Arizona Press), 669

\bibitem[18b]{18b}
Li, X. Q., Zhang, H., \& Li, Q. B. 1995, A\&A, 304, 617

\bibitem[19]{19}
Lin, D. N. C., Bodenheimer, P., \& Richardson, D. C. 1996, Nature, 380, 606

\bibitem[19b]{19b}
Lions, P. L. 1982, SIAM Rev., 24, 441

\bibitem[20]{20}
Lissauer, J. J. 1993, \araa, 31, 129

 %\bibitem[20aa]{20aa}
 %Lovis, C., Mayor, M., Pepe, F., et al. 2006, Nature, 441, 305

\bibitem[20aaa]{20aaa}
Lynch, P. 2003, \mnras, 341, 1174

\bibitem[20a]{20a}
Malhotra, R. 1993, Nature, 365, 819

\bibitem[20ab]{20ab}
Malhotra, R. 1995, \aj, 110, 420

\bibitem[21]{21}
Marcy, G. W., \& Butler, R. P. 1996, \apj, 464, L147

\bibitem[22]{22}
Marcy, G. W., Butler, R. P., Fischer, D. A., et al. 2005, 
Prog. Theor. Phys. Suppl., 158, 24

\bibitem[23]{23}
Mayor, M., \& Queloz, D. 1995, Nature, 378, 355

 %\bibitem[23b]{23b}
 %McArthur, B. E., Endl, M., Cochran, W. D., et al. 2004, \apj, 614, L81

 %\bibitem[24]{24}
 %McFadden, L. A., Tholen, D. J., \& Veeder, G. J. 1989, in Asteroids
 %II, ed. R. P. Binzel, T. Gehrels, \& M. S. Matthews (Tucson:
 %Univ. of Arizona Press), 442

\bibitem[23c]{23c}
Millan--Gabet, R., Schloerb, F. P., \& Traub, W. A. 2001, \apj, 546, 358

\bibitem[24a0]{24a0}
Morbidelli, A., Brown, M. E., \& Levison, H. F. 2003, Earth, Moon \& Planets, 92, 1

\bibitem[24b02]{24b02}
Narita, S., Kiguchi, M., Miyama, S. M., \& Hayashi, C. 1990, \mnras, 244, 349

\bibitem[24aa]{24aa}
Neuh$\ddot{\rm a}$user, R., \& Feitzinger, J. 1986, A\&A, 170, 174

\bibitem[24a0a]{24a0a}
Nieto, M. M. 1972, The Titius--Bode Law of Planetary Distances: Its History
and Theory (Oxford: Pergamon Press)

\bibitem[24a]{24a}
Nottale, L., Schumacher, G., \& Gay, J. 1997, A\&A, 322, 1018

\bibitem[24b]{24b}
Ostriker, J. 1964, \apj, 140, 1056

 %\bibitem[25]{25}
 %Press, W. H., Flannery, B. P., Teukolsky, S. A., \& Vetterling, W. T. 1988,
 %Numerical Recipes (Cambridge: Cambridge Univ. Press)

\bibitem[25a]{25a}
Rix, H-W, de Zeeuw, P. T., Cretton, N., van der Marel, R. P., \&
Carollo, C. M. 1997, \apj, 488, 702

\bibitem[25b]{25b}
Robe, H. 1968, Ann. d'Astrophys., 31, 549

\bibitem[26]{26}
Ruzmaikina, T. V., Safronov, V. S., \& Weidenschilling, S. J. 1989, in
Asteroids II, ed. R. P. Binzel, T. Gehrels, \& M. S. Matthews (Tucson:
Univ. of Arizona Press), 681

\bibitem[26a]{26a}
Sarazin, C. L. 1988, X--Ray Emission from Clusters of Galaxies
(Cambridge: Cambridge Univ. Press)

\bibitem[26p]{26p}
Schmitz, F. 1984, A\&A, 131, 309

\bibitem[26p]{26p}
Schmitz, F., \& Ebert, R. 1986, A\&A, 154, 214

\bibitem[26bc]{26bc}
Schneider, M., \& Schmitz, F. 1995, A\&A, 301, 933

\bibitem[26b]{26b}
Sornette, D. 1998, Phys. Rep., 297, 239

\bibitem[26c]{26c}
Stod\'olkiewicz, J. S. 1963, Acta Astron., 13, 30

\bibitem[27]{27}
Strom, S. E., Edwards, S., \& Skrutskie, M. F. 1993, in Protostars
and Planets III, ed. E. H. Levy, \& J. I. Lunine (Tucson: Univ.
of Arizona Press), 837

\bibitem[27a0]{27a0}
Thomas, L. H. 1927, Proc. Cambridge Phil. Soc., 23, 542

 %\bibitem[27a01]{27a01}
 %Thomson, W. 1862, Manchester Phil. Soc. Proc., 2 (170), 125

\bibitem[27a00]{27a00}
Tohline, J. E. 2002, \araa, 40, 349

\bibitem[27a]{27a}
Visser, M., \& Yunes, N. 2003, Int. J. Mod. Phys., 18, 1

\bibitem[27bc]{27bc}
Vogt, S. S., Butler, R. P., Marcy, G. W., et al. 2005, \apj, 632, 638

\bibitem[28]{28}
Weidenschilling, S. J. 1977, A\&SS, 51, 153

\bibitem[29]{29}
Wetherill, G. W. 1989, in Asteroids II, ed. R. P. Binzel, T. Gehrels,
\& M. S. Matthews (Tucson: Univ. of Arizona Press), 661

\bibitem[30]{30}
Wong, J. S. W. 1975, SIAM Rev., 17, 339

\end{thebibliography}
\end{document}